\documentclass[jgrga]{agutex}

\usepackage{graphicx}
\authorrunninghead{TATJANA ZIVKOVIC, KRISTOFFER RYPDAL}

\titlerunninghead{MAGNETOSPHERE DURING STORMS}

\authoraddr{T.  \v{Z}ivkovi\'{c},
Department of physics and Technology, University of Troms{\o},
Prestvannveien 40, 9037 Troms{\o}, Norway
(tatjana.zivkovic@uit.no)}

\authoraddr{K. Rypdal, Department of physics and Technology, University of Troms{\o},
Prestvannveien 40, 9037 Troms{\o}, Norway}

\begin{document}

%
%

\title{Low-dimensionality and predictability of solar wind and  global  magnetosphere during magnetic storms}
\authors{T.  \v{Z}ivkovi\'{c}, \altaffilmark{1}
and K. Rypdal, \altaffilmark{1}}

\altaffiltext{1}{Department of Physics and Technology,
University of Troms{\o}, Norway.}

\begin{abstract}
The storm index SYM-H, the solar wind velocity $v$, and  interplanetary magnetic field $B_{z}$ show no signatures of low-dimensional dynamics in quiet periods, but tests for determinism in the time series indicate that SYM-H exhibits a significant  low-dimensional component during storm time, suggesting that self-organization takes place during magnetic storms. Even though our analysis yields no discernible change in determinism during magnetic storms for the solar wind parameters, there are significant enhancement of the predictability and exponents measuring persistence. Thus, magnetic storms are typically  preceded by an increase in the persistence of the solar wind dynamics, and this increase is also present in the magnetospheric response to the solar wind.

\end{abstract}

\begin{article}

%
%

\section{Introduction}
Under the influence of the solar wind, the magnetosphere resides in a complex, non-equilibrium state. 
The plasma particles have non-Maxwellian velocity distribution, MHD turbulence is present everywhere, and  intermittent energy transport known as bursty-bulk flows occurs as well \citep{A99}. The magnetospheric response to particular  solar events constitutes an essential aspect of space weather
 while the  response to solar variability in general is often referred to as {\it space climate} \citep{w2002}.    Theoretical approaches to space climate involve  concepts and methods  from stochastic processes, nonlinear dynamics and  chaos, turbulence, self-organized criticality, and phase transitions.

Self-organization  can lead to low-dimensional behavior in the magnetosphere \citep{K96,V90,S93}. However, power-law dependence observed in the Fourier spectra of the auroral electrojet (AE) index is a typical signature of high dimensional colored noise indicating  multi-scale dynamics of the magnetosphere. 
In order to reconcile low-dimensional, deterministic behavior with  high-dimensionality, \cite{T98}  proposed that a high-dimensional system near self-organized criticality (SOC) \citep{B87} can be characterized by a few parameters whose  evolution is  governed by a small number of nonlinear equations. Some magnetospheric models, like the one presented in \cite{c98}, are based on the SOC-concept. Here a system tunes itself to criticality and the energy transport  across scales is mediated by avalanches which are power-law distributed in size and duration. 

On the other hand, it was suggested in \cite{S2001}  that the substorm dynamics can be described as a non-equilibrium phase transition; i.e.  as a system tuned externally to criticality.  Here, a power-law relation is given, with characteristic exponent close to the input-output critical exponent in a second-order phase transition. In fact, it is claimed in \cite{Sh03} that the global features of the magnetosphere correspond to a first order phase transition whereas multi-scale processes correspond to the second- order phase transitions.

The existence of  metastable states in the magnetosphere, where intermittent signatures might be due to dynamical phase transitions among these states, was suggested by \cite{CG2001}, and forced and/or self-organized criticality (FSOC) induced by the solar wind was introduced as a conceptual description of magnetospheric dynamics. 
The concept of intermittent criticality was suggested by \cite{B06} who asserted  that during intense magnetic storms the system develops long-range correlations, which further indicates a transition from a less orderly to a more orderly state. Here, substorms might be the agents by which longer correlations are established. This concept implies a time-dependent variation in the activity as the critical point is approached, in contrast to SOC. 

In the present paper we investigate determinism  and predictability of observables characterizing the state of the magnetosphere during geomagnetic storms as well as during its quiet condition, but the emphasis is on the evolution of these properties over the course of major magnetic storms. The measure of determinism employed here increases if the system dynamics is dominated by modes governed by low-dimensional dynamics. Hence, the determinism in most cases is a measure of low-dimensionality.  For a low-dimensional, chaotic system the predictability measure increases when the largest Lyapunov exponent increases, and hence it is really a measure of un-predictability. For a high-dimensional or stochastic system it is related to the degree of persistence in time series representing the dynamics. High persistence means high predictability.

One of the most useful data tools for probing the magnetosphere during substorm conditions is the AE minute index which is defined as the difference between the AU index, which measures the eastward electrojet  current in the auroral zone, and the AL index, which measures the westward electrojet current, and is usually derived from 12 magnetometers  positioned under the auroral oval \citep{DS66}. The auroral electrojet, however, does not respond strongly to the specific modifications of the magnetosphere that occur during magnetic storms. A typical storm characteristic, however,  is a change in the intensity of the symmetric part of the ring current that encircles Earth at altitudes ranging from about 3 to 8 Earth radii, and is proportional to the total energy in the drifting particles that form this current system \citep{G94}.  The indices $D_{st}$ and SYM-H indices  are both designed  for the study of storm dynamics. These indices  contain contribution from the magnetopause current, the partial and symmetric ring current, the substorm current wedge, the magnetotail currents, and induced currents on the Earth's surface. They are derived from similar data sources, but SYM-H has the distinct advantage of having 1-min time resolution compared to the 1-hour time resolution of $D_{st}$. \cite{Wan06} has recommended that the SYM-H index be used as a de facto high-resolution $D_{st}$ index. The analysis of these indices are central to this study. We particularly focus on SYM-H and SYM-H$^{\star}$ which is derived from the SYM-H when the contribution of the magnetopause current is excluded.

The typical magnetic storm consists of the initial phase, when the horizontal magnetic field suddenly increases and stays elevated for several hours, the main phase where this component is depressed for one to several hours, and the recovery phase which
also lasts several hours. The initial phase has been  associated with northward directed IMF (little energy enters the magnetosphere), but it has been discovered that this phase is not essential for the storm to occur \cite{A65}.
In order to define a storm, we follow the approach of  \cite{LP97}, where the $D_{st}$ minimum is a common reference epoch, the main-phase decrease is sufficiently steep, and the recovery phase is also defined. 

\subsection{Data acquisition}

The SYM-H index data are downloaded from World Data Center, with 1-min resolution. We also use minute data for the interplanetary magnetic field (IMF) component $B_{z}$,  minute data for the solar wind bulk velocity $v$ along the
Sun-Earth axis, as well as flow pressure which is given in nT. These data are retrieved from the OMNI satellite database and are given in the GSE coordinate system. 
Gaps of missing data in $B_{z}$, $v$ and flow pressure  are linearly interpolated from the data which are not missing, while SYM-H data are analyzed for the
entire period. The same result for the  $B_{z}$ and $v$ is obtained when gaps of missing data are excluded from the analysis. 

Data for the period from January 2000  till December 2005 is used to compute general properties of the magnetosphere. In order to analyze storm conditions all the indices are analyzed during ten intense magnetic storms. Analyzed storms occurred on 6 April 2000, 15 July 2000, 12 August 2000, 31 March 2001, 21 October 2001, 28 October 2001, 6 November 2001, 7 September 2002, 29 October 2003, and 20 November  2003. These storms are characterized with $D_{st}$ minimum which is in the range between -150 nT to -422 nT. \\

The remainder of the paper is organized as follows: section 2 describes the data analysis methods employed. Section 3 presents analysis results discerning general statistical scaling properties of global magnetospheric dynamics using minute data over several years and data generated by a numerical  model which produces realizations of a fractional Ornstein-Uhlenbeck (fO-U) process. In particular we study how determinism and predictability of the geomagnetic and solar wind observables change over the course of magnetic storms. Section 4 is reserved for  discussion of results and  section 5 for conclusions.

\section{Methods}
\subsection{Recurrence-plot analysis}

The recurrence plot is a powerful tool for  the visualization of recurrences of  phase-space trajectories.
It is very useful since it can  be applied to  non-stationary as well as short time series  \citep{EK}, and this is  the nature of  data we use to explore magnetic storms. Prior to constructing a  recurrence plot the common procedure is to reconstruct phase space  from the time-series $x(t)$ of length $N$ by time-delay embedding \citep{T81}.

Suppose the physical  system at hand is a deterministic dynamical system describing the evolution of a state vector ${\bf z}(t)$ in a phase space of dimension $p$, i.e. $\bf z$ evolves according to an autonomous system of 1st  order ordinary differential equations;
\begin{equation}
 \frac{d{\bf z}}{dt}={\bf f}({\bf z}),\, \, \, {\bf f}: {\cal R}^p\rightarrow {\cal R}^p \label{eq1}
 \end{equation}
and that an observed time series $x(t)$ is generated by the measurement function $g: {\cal R}^p\rightarrow {\cal R}$,
\begin{equation}
x(t)=g({\bf z}(t)). \label{eq2}
\end{equation}
Assume that the dynamics takes place on an invariant set (an attractor) ${\cal A} \subseteq {\cal R}^p$ in phase space, and that this set has box-counting fractal  dimension $d$. Since the dynamical system  uniquely defines the entire phase-space trajectory once the state ${\bf z}(t)$ at a particular  time $t$ is given, we can define uniquely an $m$-dimensional measurement function, 
\begin{equation}
{\bf g}: {\cal A}\rightarrow {\cal R}^m, \, \, {\bf g}({\bf z})=(x(t),x(t+\tau),\ldots,x(t+(m-1)\tau)). \label{eq3}
\end{equation}
 where  the vector components are given by equation (\ref{eq2}), and $\tau$ is a  time delay of our choice. If the invariant set $\cal A$ is compact (closed and bounded),  $g$ is a smooth function and $m>2d$, the map given by equation (\ref{eq3}) is a topological embedding (a one-to-one  continuous map) between $\cal A$ and ${\cal R}^m$. The condition $m>2d$ can be thought of as a condition for the image ${\bf g}({\cal A})$ not to intersect itself, i.e. to avoid that two different states on the attractor $\cal A$ are mapped to the same point in the $m$-dimensional embedding space ${\cal R}^m$. If such an   embedding is  achieved, the trajectory ${\bf x}(t)={\bf g}({\bf z})$ (where ${\bf g}({\bf z})$ is given by equation (\ref{eq3})) in the embedding space is a complete mathematical representation of the dynamics on the attractor. Note that the dimension $p$ of the original phase space is irrelevant for the reconstruction of the embedding space. The important thing is the dimension $d$ of the invariant set $\cal A$ on which the dynamics unfolds.

There are practical constraints on useful choices of the time delay $\tau$. If $\tau$ is much smaller than the autocorrelation time the  image of $\cal A$ becomes  essentially one-dimensional. If $\tau$ is much larger than the autocorrelation time, noise may destroy the deterministic connection between the components of  ${\bf x}(t)$, such that our assumption that ${\bf z}(t)$ determines ${\bf x}(t)$ will fail in practice. A common choice of $\tau$ has been the first minimum of the autocorrelation function, but it has been shown that better results are achieved by selecting the
 time delay  as the first minimum in the average mutual information function, which can be percieved as a nonlinear autocorrelation function \citep{DA}. Here we use the average mutual information function to calculate the value of $\tau$.
 
 The recurrence-plot analysis deals with the trajectories in the embedding space. If the original time series $x(t)$ has $N$ elements, we have a time series of $N-(m-1)\tau$ vectors ${\bf x}(t)$ for $t=1,2,\ldots,N-(m-1)\tau$. This time series constitutes the trajectory in the reconstructed embedding space.

The next step is to construct a $[N-(m-1)\tau]\times [(N-(m-1)\tau]$ matrix $R_{i,j}$ consisting of elements 0 and 1. The matrix element $(i,j)$ is 1 if  the distance is $\Vert {\bf x}_i-{\bf x}_j\Vert \leq \epsilon$ in the reconstructed  space, and otherwise it is 0. 
The recurrence plot is simply a plot where the points $(i,j)$ for which the corresponding matrix element is 1 is marked by a dot. For a  deterministic system the radius $\epsilon$ is typically  chosen as 10\% of the diameter of the reconstructed attractor, but varies for different sets of data. For a non-stationary stochastic process like a Brownian motion there is no bounded attractor for the dynamics, and the diameter is limited by the length of the data record.
The first example of recurrence plot is shown in Figure 1,  obtained from  the $B_{z}$ when no storm is present on 5 September 2001. In Figure 2, the recurrence plot is shown for the $B_{z}$ for the strong storm on 6 April 2000. In both cases, embedding dimension is $m=1$ and
$\epsilon \sim 0.4$, which corresponds to 10\% of the data range.
 
 \begin{figure} \label{fig1}
\begin{center}
\includegraphics[width=8cm]{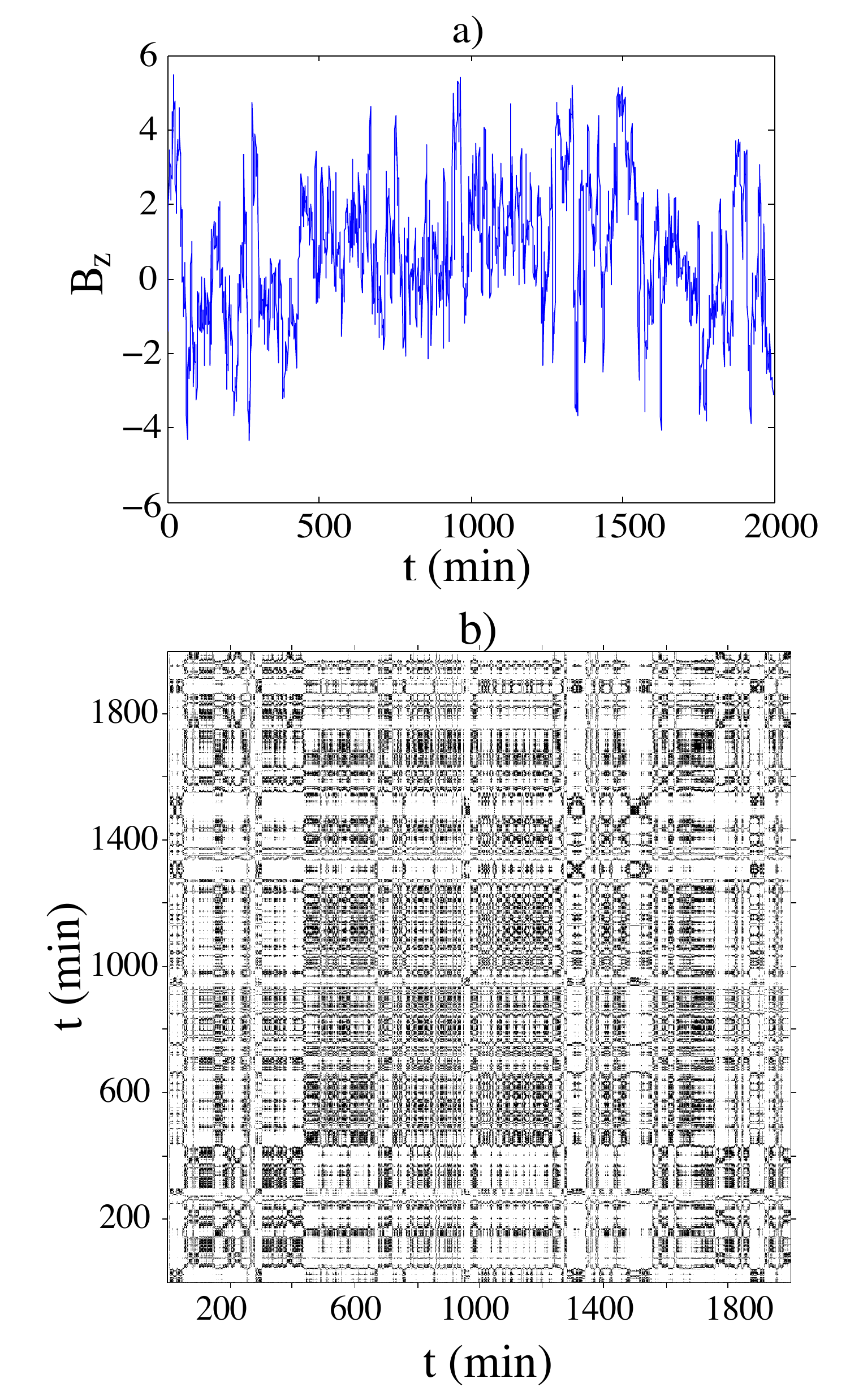} 
\caption{$B_{z}$ during quiet condition on September 5, 2001. a) $B_z$ time series, b) Recurrence plot of the time  series shown  in (a).}
\end{center}
\end{figure}

 \begin{figure} \label{fig2}
\begin{center}
\includegraphics[width=8cm]{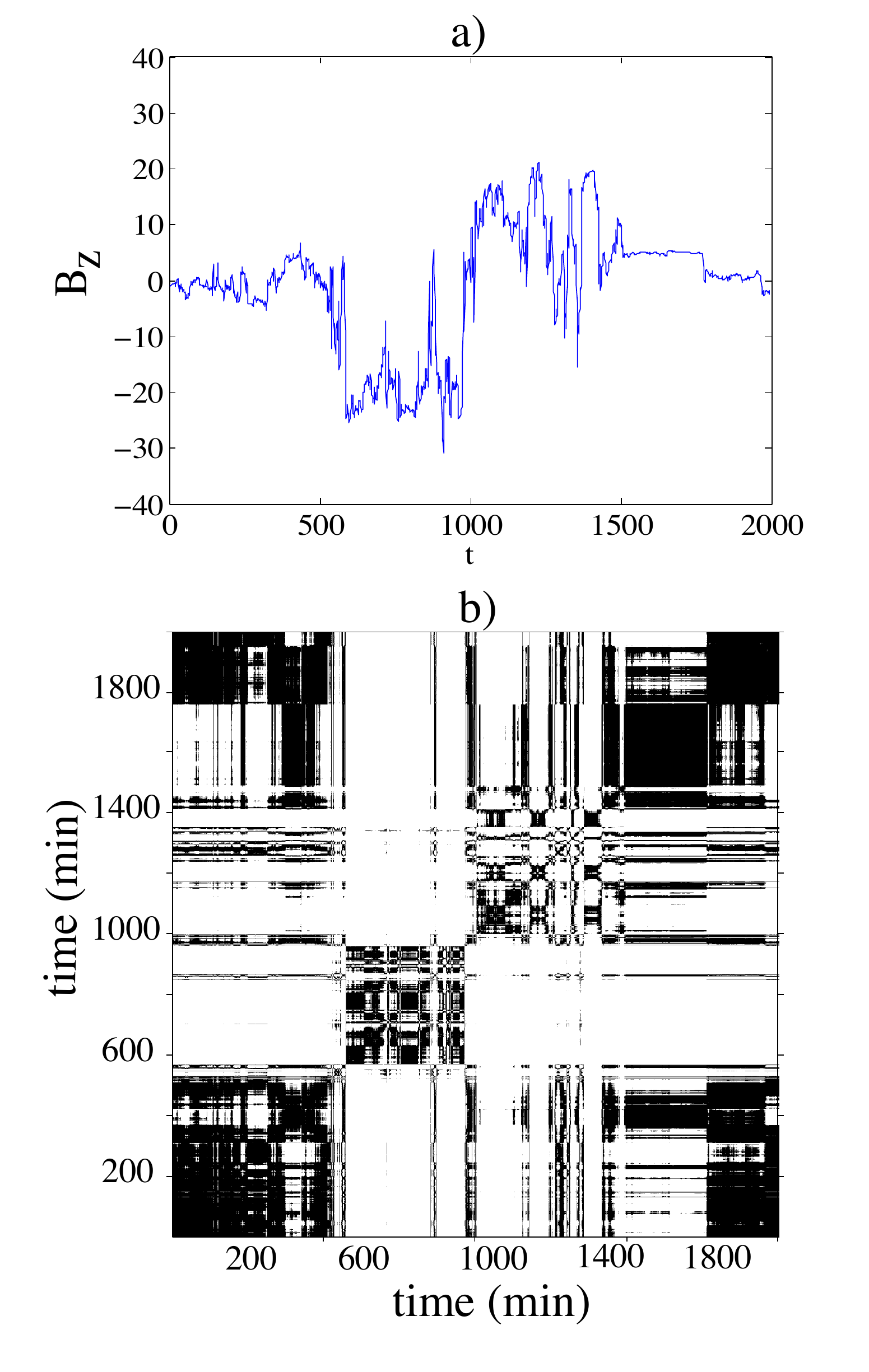} 
\caption{$B_{z}$ during the strong storm on 6 April 2000. a)  $B_z$ time series, b) Recurrence plot of the time  series shown  in (a).}
\end{center}
\end{figure}

\subsection{Empirical mode decomposition}\label{EMDmethod}
The empirical mode decomposition (EMD) method, developed in \cite{H98} is very useful on non-stationary and nonlinear time series. 
EMD method can give a change of frequency in any moment of time (instantaneous frequency) and a change of amplitude in the system.
However, in order to properly define instantaneous frequency, a time series should have the same number of zero crossings and extrema (or they can differ at most by one), and
a local mean should be close to zero. The original time series usually does not have these characteristics and should be decomposed into intrinsic mode functions (IF) for which
instantaneous frequency can be defined.
Decomposition can be obtained through the so-called  sifting process. This is an adaptive process derived from the data and
can be briefly described as follows:  All local maxima and minima in the time series $s(t)$ are found, and all local
maxima and minima are fitted by cubic spline and these fits define the upper (lower) envelope of the time series. Then the mean of the upper and lower envelope $m(t)$ is defined, and the difference between the time series and this mean represents the first IF, $h(t)=s(t)-m(t)$, if instantaneous frequency can be obtained, defined by some stopping criterion. If not, the  procedure is repeated (now starting from $h(t)$ instead of $s(t)$) until the first IF is produced.
Higher IFs are obtained by subtracting the first IF  from the time series $s(t)$ and the entire previously mentioned procedure 
is repeated until a residual, usually a monotonic function, is left. We use a stopping criterion defined by \cite{R03}, where $\eta(t)<\theta_{1}$ on $1-\gamma$ fraction of the IF, and $\eta(t)<\theta_{2}$, on the remaining fraction of the IF. Here $\eta=m(t)/a(t)$, $a(t)$ is the IF amplitude, and $\gamma=0.05$, $\theta_{1}=0.05$, and $\theta_{2}=0.5$. By the above definitions, IFs
are complete in the sense that their summation gives the original time series: $s(t)=\sum_{1}^{M} h(t)+R(t)$ where $M$ is the number of IFs and $R$ is a residual. In Figure 3a we show the IFs from EMD performed on the IMF $B_{z}$ during a magnetic storm on 6  April  2000 (whose time series is plotted in Figure 2a), while in Figure 3b the $D_{st}$ index for the same storm is shown.
 
 \begin{figure} \label{fig3}
\begin{center}
\includegraphics[width=8cm]{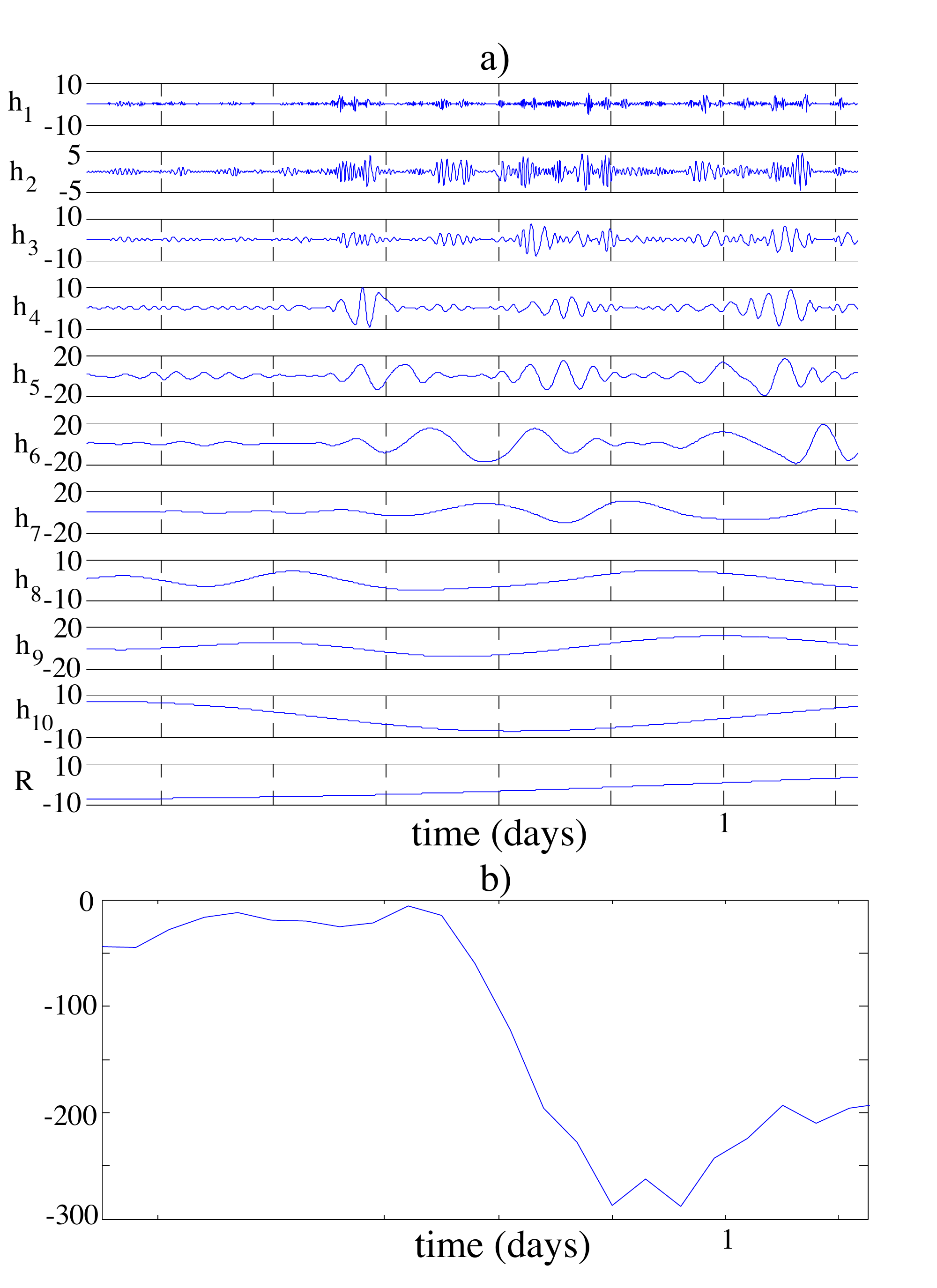} 
\caption{a) Intrinsic mode functions obtained by EMD for $B_{z}$ for the magnetic storm on 6 April 2000, b) $D_{st}$ for the same event.}
\end{center}
\end{figure}

In order to study  stochastic behavior of a time series by means of EMD analysis, we refer to \cite{WH2004} who studied characteristics of white noise using the EMD method. They derived for white noise  the relationship $\log E_{m}=-\log T_{m}$, where $E_{m}$ and $T_{m}$ represents empirical variance and mean period for the $m$'th IF. 
Here, $E_{m}=(1/N)\sum_{t=1}^{N} h(t)^{2}$, where $h(t)$ is the {\it m}'th IF and $T_{m}$ is the ratio of the {\it m}'th IF length to the number of its zero crossings.
\cite{F2009} analyzed telecommunication indices and noticed  a resemblance to autoregressive processes of the first order AR(1), which are stochastic and linear processes. For such processes  $\log\ E_{m}=\zeta \log T_{m}$. For fractional Gaussian noise processes ($H<1$) and fractional Brownian motions ($H>1$) we have the connection $\zeta=2H-2$, where $H$ is the Hurst exponent, as shown by \cite{FG04}. 
A useful feature of the EMD analysis is the possibility of extraction of trends in the time series \citep{W2007}, because the slowest IF components should often be interpreted  as trends. This is an advantage compared to the standard variogram or rescaled-range techniques \citep{Beran}, whose estimation of the scaling exponents is biased by the trend.
\subsection{A test for determinism}\label{determinism}
In this paper we employ a simple test for determinism, developed by \cite{KG92}, where the following hypothesis is tested: When a system is deterministic, the orientation of the trajectory (its tangent) is a function of the position in the phase space. Further, this means that the tangent vectors of a  trajectory which recurs to  the same small ``box'' in  phase space, will have the same  directions since these are uniquely determined by the position in  phase space. On the other hand, trajectories in a stochastic system have directions which do not depend uniquely on the position and are equally probable in any direction. This test works only for continuous flows, and is not applicable to maps since consecutive points on the orbit may be very separated in the phase space.
For flows, the trajectory orientation is defined by a vector  of a unit length, whose direction is given by the displacement between the point where trajectory enters the box {\it j} to the point where the trajectory exits the same box.
The displacement in {\it m}-dimensional embedding space is given from the time-delay embedding reconstruction: 
\begin{eqnarray}
\Delta {\bf x}(t)&=&[ x(t+b)-x(t), x(t+\tau+b)-x(t+\tau),\ldots , \nonumber  \\ 
& &x(t+(m-1)\tau+b)-x(t+(m-1)\tau)] , \label{eq6}
\end{eqnarray}
 where $b$ is the time the trajectory spends inside a box.
The orientation vector for the $k$th pass through box $j$ is the unit vector ${\bf u}_{k,j}=\Delta {\bf x}_{k,j}(t)/\vert \Delta {\bf x}_{k,j}(t)\vert$.
The  estimated averaged displacement  vector in the box is
 \begin{equation}
{\bf V}_{j}=\frac{1}{n_{j}}\sum_{k=1}^{n_{j}}{\bf u}_{k,j}, \label{eq7}
\end{equation}
 where $n_j$ is the number of passes of the trajectory through box $j$.  If the dynamics is deterministic,  the embedding dimension is sufficiently high, and in the limit of vanishingly small box size, the trajectory directions should be aligned and $V_j\equiv \vert  {\bf V}_{j}\vert=1$. In the case of  finite box size, ${\bf V}_{j}$ will not depend very much on the  number of passes $n_j$, and $V_j$ will  converge to $1$  as $n_j\rightarrow \infty$. In contrast, for the trajectory of a random process, where the direction of the next step is completely independent of the past, $V_j$ will decrease with $n_j$ as $V_j\sim n_j^{-1/2}$. In our analysis we will choose the linear box dimension equal to the mean distance a phase-space point moves in one time step and set $b=1$ time step in equation (\ref{eq6}).
 
 \begin{figure} \label{fig4}
\begin{center}
\includegraphics[width=6cm]{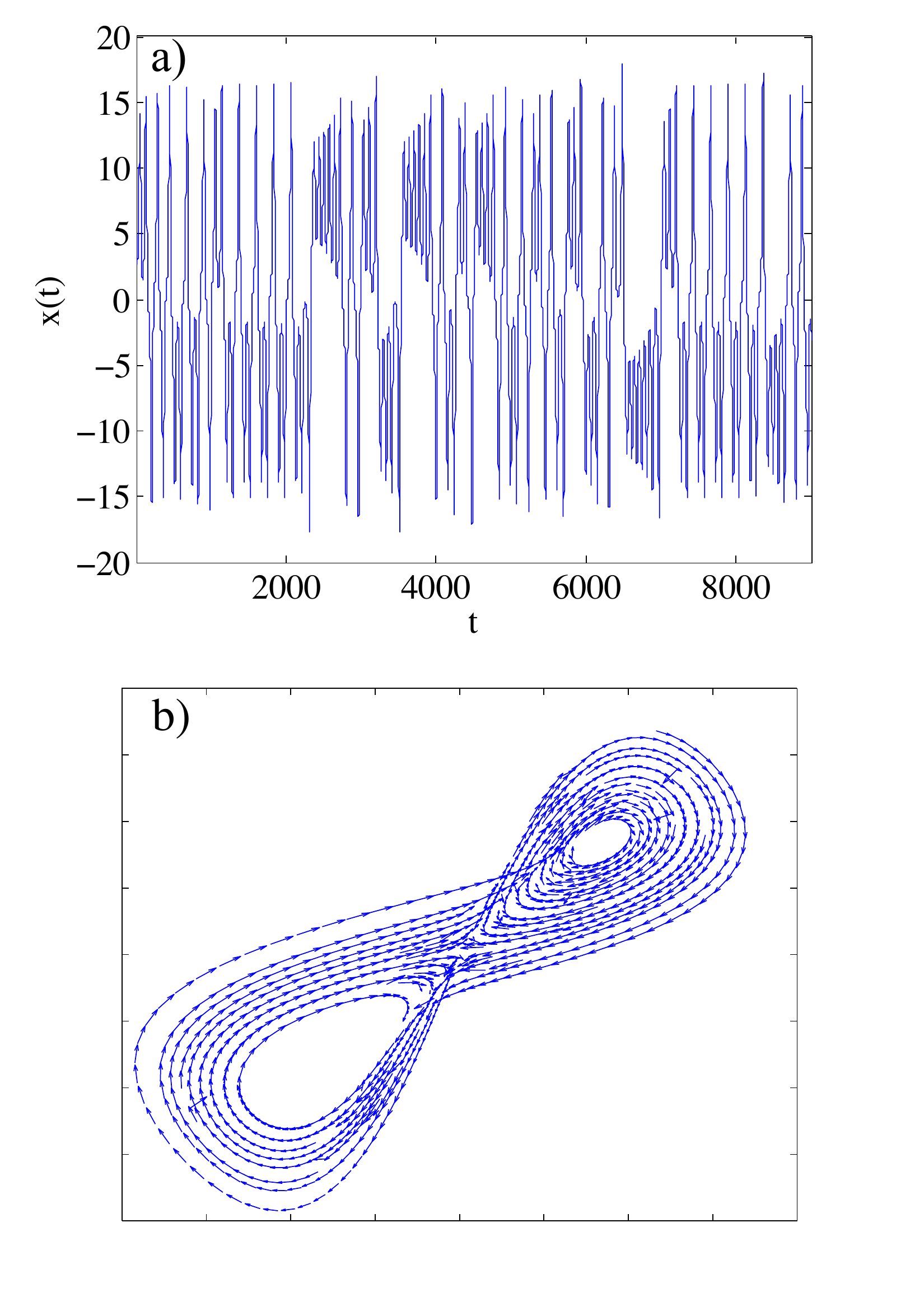} 
\caption{ a) Time series representing one component of the numerical solution of the Lorenz system. b) Average displacement vectors ${\bf V}_j$ in each  box visited by  a 2-dimensional projection of  the $m=3$ -dimensional embedding  space reconstructed from the time series in (a).}
\end{center}
\end{figure}

 \begin{figure} \label{fig5}
\begin{center}
\includegraphics[width=6cm]{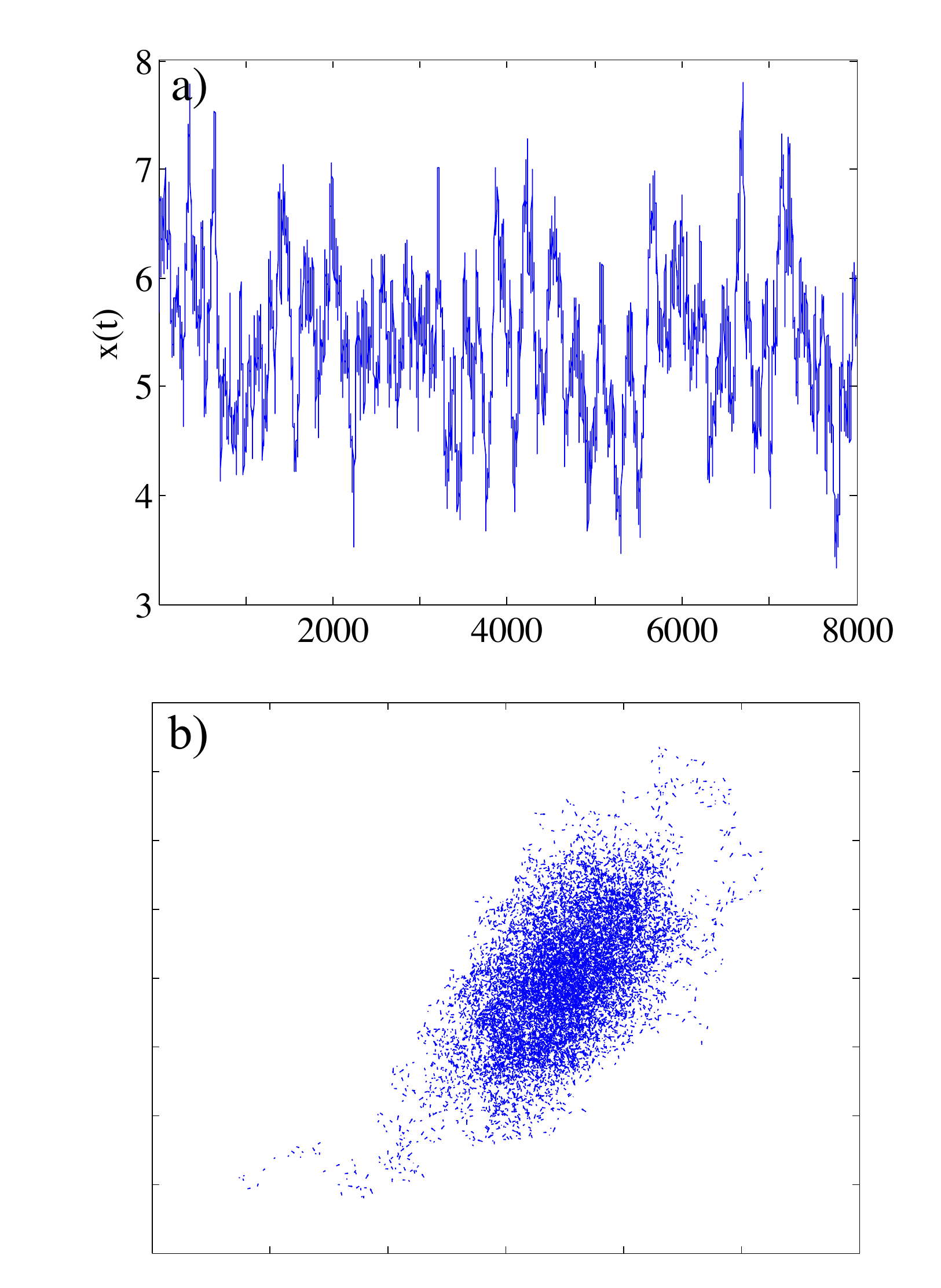} 
\caption{a) Time series representing the numerical solution of the equation for the f-OU process. b) Average displacement vectors ${\bf V}_j$ in each  box visited by  a 2-dimensional projection of  the $m=8$ -dimensional embedding  space reconstructed from the time series in (a).}
\end{center}
\end{figure}

In Figure  4b we show  displacement vectors  ${\bf  V}_{j}$ averaged over the passes through the box $j$, for a three-dimensional embedding of  the Lorenz attractor, whose time series is shown in Figure 4a; in Figure 5b the same is shown for a random process, in this case a fractional Ornstein-Uhlenbeck (fO-U) process. These model systems will be used throughout this paper as archetypes of low-dimensional and stochastic systems, respectively. The Lorenz system has the form
\begin{eqnarray} \label{eq13}
      dx/dt&=&a (y-x)      \\
      dy/dt&=&-x z+c x-y            \nonumber \\
      dz/dt&=&x y-b z         \nonumber,
\end{eqnarray}
with standard coefficient values  $a=10$, $b=8/3$, and $c=28$, which give rise to  a chaotic flow.
The fO-U process is described  by the stochastic equation:
\begin{equation}
dS_{t}=\lambda (\mu-S_{t})+\sigma dW_{t},
\end{equation}
where $dW_{t}$ is a fractional Gaussian noise with Hurst exponent $H$ \citep{Beran}.
The drift ($\lambda$ and $\mu$) and diffusion ($\sigma$) parameters are fitted by the least square regression to  the time  series of the SYM-H storm index. This will be explained in more detail in section \ref{stormdeterminism}.

The  degree of determinism of the dynamics can be assessed by exploring the dependence of $ V_j$ on $n_j$. In practice, this can be done by computation of  the averaged displacement vector
 \begin{equation}
 L_n\equiv\langle  V_{j}  \rangle_{n_j=n}, \label{eq8}
 \end{equation} where the average is done over all boxes with same number $n$ of trajectory passes. As shown in \cite{KG93}, the average displacement of $n$ passes in {\it m}-dimensional phase space for the Brownian motion is \begin{equation}
R_n=\frac{1}{\sqrt{n}}(\frac{2}{m})^{1/2}\frac{\Gamma\lbrack (m+1)/2\rbrack}{\Gamma(m/2)},\label{eq9}
\end{equation} 
 where $\Gamma$ is the  gamma function.
The deviation in $\langle V_{j}\rangle$  between a given time series and the Brownian motion can be characterized by a single number given by the weighted average over all boxes of the quantity,
\begin{equation}
\Lambda(\tau)\equiv \frac{1}{\sum_{j} n_{j}}\sum_{j} n_{j} \frac{\langle V_j\rangle ^{2}(\tau)-R_{n_j}^{2} }{1-R_{n_j}^{2}},\label{eq10}
\end{equation}
where we have explicitly highlighted that the averaged displacement $\langle V_j\rangle(\tau)$ of the trajectory in the reconstructed phase space depends on the time-lag $\tau$.   For a completely deterministic signal we have $\Lambda=1$, and for a completely random signal $\Lambda =0$.

All systems described by the laws of classical (non-quantum) physics are deterministic in the sense that they are described by equations that have unique solutions if the initial state is completely specified. In this sense it seems meaningless to provide tests for determinism. The test described in this section is really a test of {\em low dimensionality}. The test is performed by means of  a time-delay embedding, for embedding dimension $m$ up to a maximum value $M$, where $M$ is limited by practical constraints. High $M$ requires longer time series in order to achieve adequate statistics. A test that fails to characterize the system as deterministic for  $m\leq M$ in reality only tells us that the embedding dimension is too small, i.e. the number of degrees of freedom $d$ of the system exceeds $M/2$. Such systems will in the following be characterized as random, or stochastic. 
 
\begin{figure} \label{fig6}
\begin{center}
\includegraphics[width=8cm]{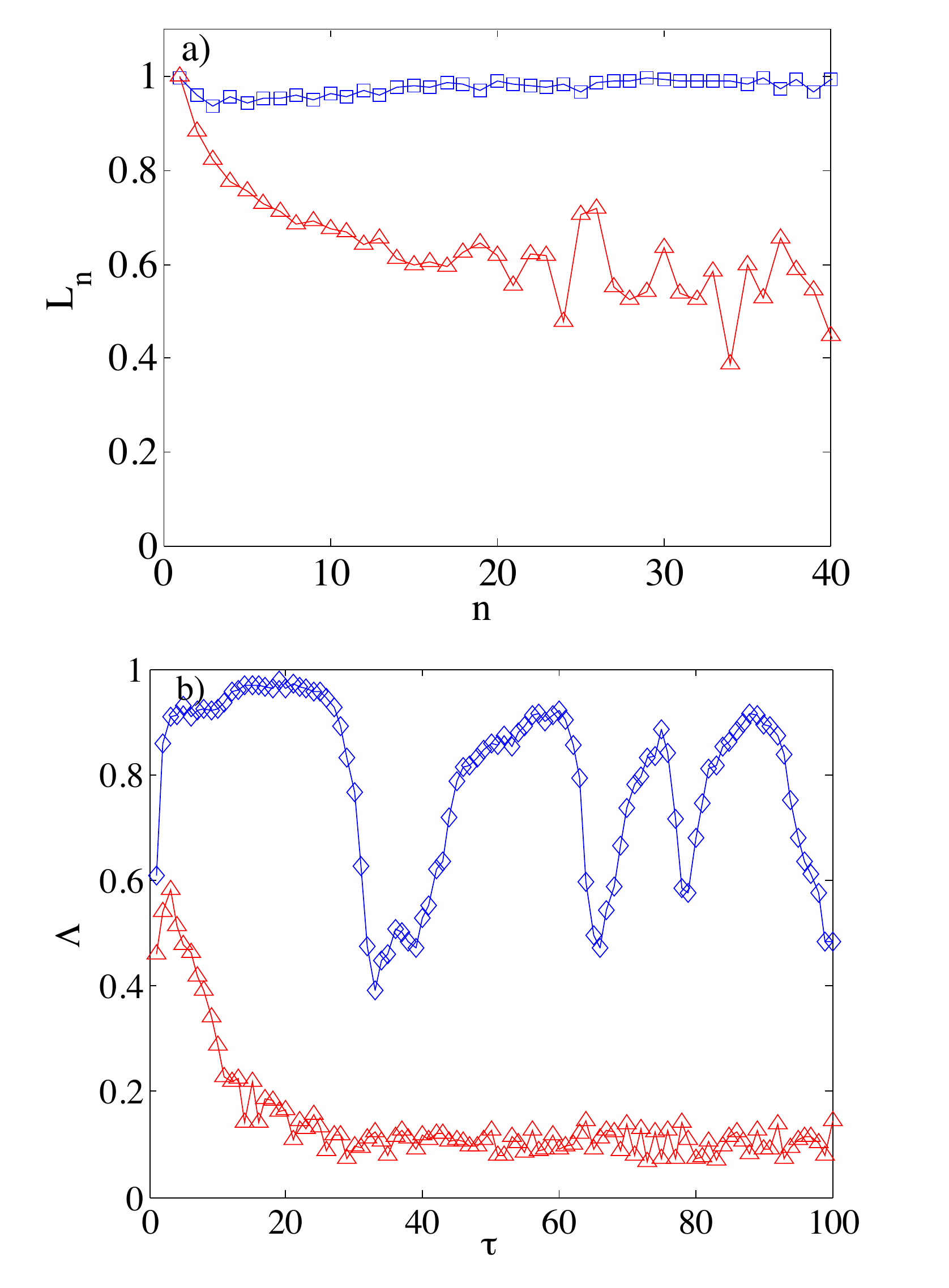}
\caption{a) $L_n$: square symbols are derived from numerical solutions of  the Lorenz system,  and  triangles from these solutions after randomization of phases of Fourier coefficients. b) $\Lambda(\tau)$: diamonds from Lorenz system, and triangles after randomization of  phases. }
\end{center}
\end{figure}

In Figure 6a, we plot $L_n$ versus $n$  for a time series generated as a numerical solution of  the Lorenz system. Here we use $m=3$, $b=1$ , $\tau=14$ and the box size is of the order of average distance a phase-space point moves during one time-step.
In the same plot we also show the same characteristic for the surrogate time series generated by randomizing the phases of the Fourier coefficients of the original time series. This procedure does not change the power spectrum or auto-covariance, but  destroys correlation between phases due to nonlinear dynamics. For low-dimensional, nonlinear systems such randomization will change $L_n$, as is demonstrated for the Lorenz system in Figure 6a. 
We also calculate $\Lambda$ versus $\tau$ for  these time series and plot the results in Figure 6b. 
Again, $\Lambda(\tau)$ for the original and surrogate time series are significantly different.

\begin{figure} \label{fig7}
\begin{center}
\includegraphics[width=8cm]{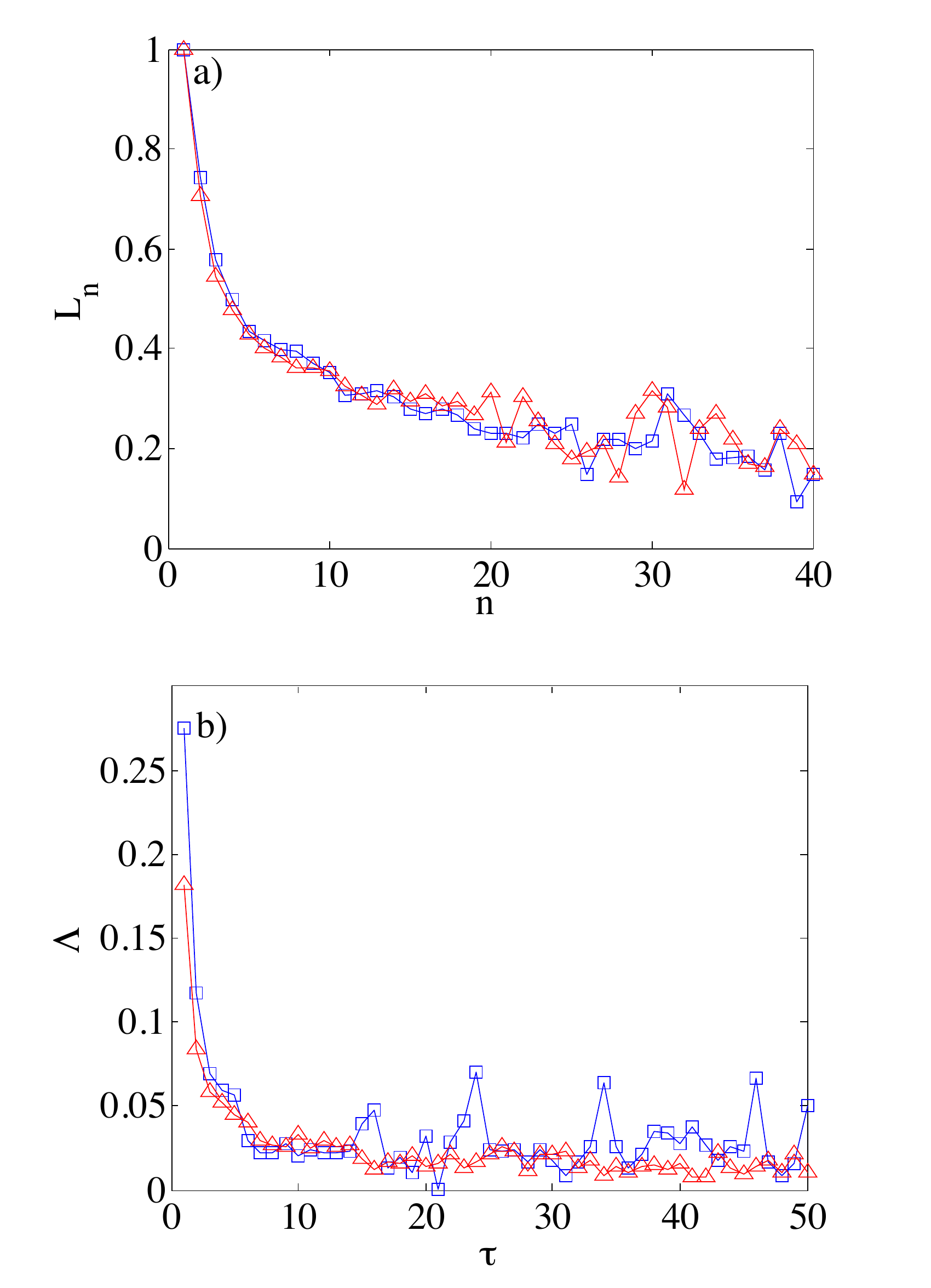}
\caption{ a) $L_n$: square symbols are derived from numerical solutions of  the fO-U stochastic equation,  and triangles from these solutions after randomization of phases of Fourier coefficients. b) $\Lambda(\tau)$: squares from fO-U equation, and triangles after randomization of  phases. }
\end{center}
\end{figure}
\begin{figure} \label{fig8}
\begin{center}
\includegraphics[width=8cm]{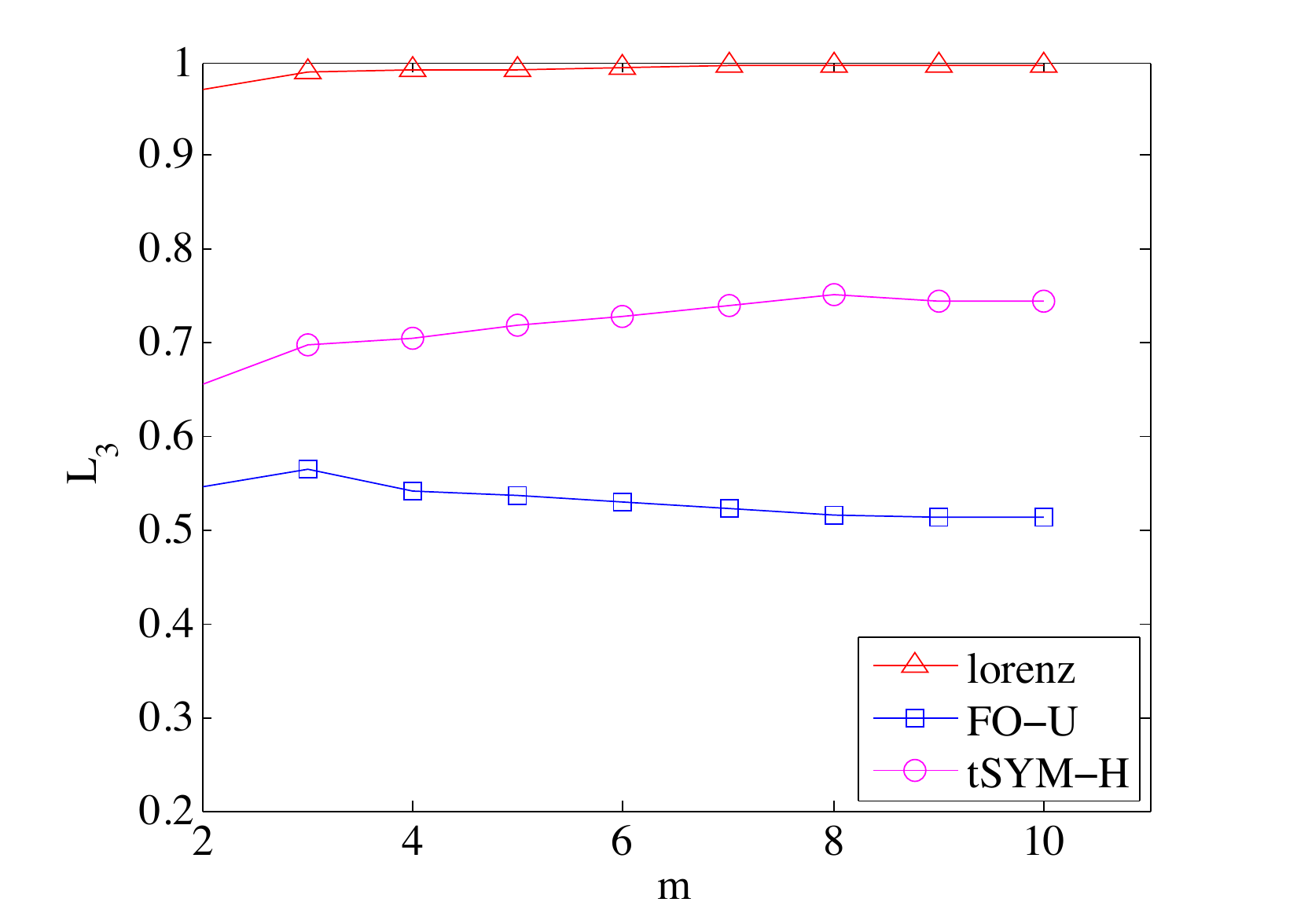}
\caption{$L_{3}$ as a function of embedding dimension $m$ for solution of Lorenz equations (triangles), fO-U process (squares), and tSYM-H (circles). }
\end{center}
\end{figure}
For the numerically generated fO-U process, where  $m=8$, $b=1$ and $\tau=20$, we observe in Figure 7 that $L_n$ and $\Lambda(\tau)$ for the original and surrogate time series do not differ, 
demonstrating that these quantities are insensitive to randomization of phases of Fourier coefficients if the process is generated by a linear stochastic equation.

One should pay attention to the nature of the experimental data used in the test of determinism. For low-dimensional data contaminated by low-amplitude noise or Brownian motions, the analysis results will  depend on the box size, but the problem is solved by choosing it sufficiently large. For a low-dimensional system represented by an  attractor of dimension $d$ the results may also depend on the choice of embedding dimension $m$. The estimated determinism $L_n$ tends to increase with increasing $m$ until it  stabilizes at $L_n\approx 1$ as $m$ approaches $2d$. For a random signal there is no such dependence on embedding dimension, as  demonstrated by example in Figure 8. Here we plot the determinism $L_3$ ($L_{n}$ when $n=3$) versus embedding dimension $m$ for the Lorenz and fO-U time series. For comparison we also plot this for transformed SYM-H, tSYM-H, during magnetic storm times (the transformation and reasons for it are explained in section \ref{sec:results}). 
It converges to a value less than 1, and for embedding dimensions higher than for the Lorenz time series. This indicates that this geomagnetic index during magnetic storms exhibit both a random and a deterministic component, and that the dimensionality of this component is higher than for the Lorenz  system.

\subsection{A test for predictability}
In this subsection we develop an analysis which is based on the diagonal line structures of the recurrence plot. In our study we use the average inverse diagonal line length:
\begin{equation}
\Gamma \equiv \langle l^{-1} \rangle=\sum_{l} l^{-1}P(l)/\sum_{l} P(l), \label{eq4}
\end{equation}
where $P(l)$ is a histogram over diagonal lengths:
$$
P(l)=\sum_{i,j=1}^{N} (1-R_{i-1, j-1} ) (1-R_{i+l, j+l}) \prod_{k=0}^{l-1} R_{i+k, j+k}.  
$$

For a low-dimensional, chaotic deterministic system (for which the embedding dimension is sufficient to unfold the attractor) $\Gamma$ is an analog to the largest Lyapunov exponent, and is a measure of the {\em  degree of unpredictability}.

For stochastic systems, the recurrence plots do not have identifiable diagonal lines, but rather consists of a pattern of dark rectangles of varying size, as observed in Figure 1. For embedding dimension $m=1$ such a dark rectangle corresponds to time intervals $I_1=(t_1,t_1+\Delta t_1)$ on the horizontal axis and $I_2= (t_2,t_2+\Delta t_2)$ on the vertical axis,  for which the signal $x(t)$ is inside the same $\epsilon$-interval whenever $t$ is included in either $I_1$ or $I_2$. In this  case the length of unbroken diagonal lines $l$ is a  characteristic measure of the linear size of the corresponding rectangle, and the PDF $P(l)$ a measure of the distribution of residence times of the trajectory inside $\epsilon$-intervals. For selfsimilar stochastic processes such as fractional Brownian motions $P(l)$ can be computed analytically, and $\Gamma$ computed as function of the selfsmilarity exponent $h$. Since the residence time $l$ inside an $\epsilon$-box increases as the smoothness of the trajectory increases (increasing $h$),  we should find that $\Gamma(h)$ is a monotonically decreasing function of $h$. In section  \ref{Sec:predictabilityresults} we compute $\Gamma(h)$ numerically for a synthetically generated fO-U process and thus demonstrate this relationship between $\Gamma$ and $h$. Hence both $\Gamma$ and $h$ can serve as measures of predictability, but $\Gamma$ is more general, because it is not restricted to selfsimilar processes or processes with stationary increments, and applies to low-dimensional chaotic as well as stochastic systems.
\begin{figure} \label{fig9}
\begin{center}
 \includegraphics[width=8cm]{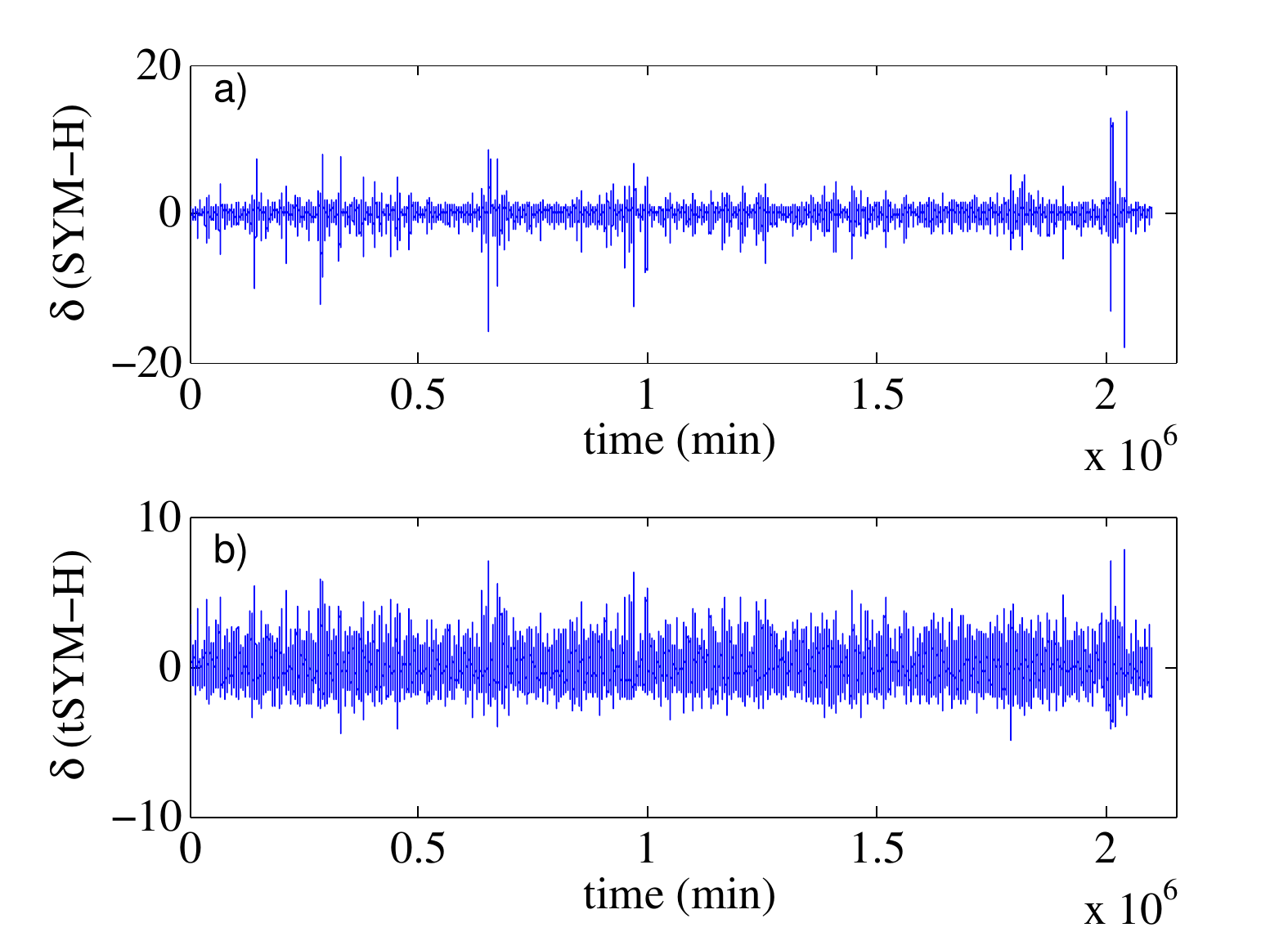}
 \caption{a) Increments for the SYM-H index. b) Increments for the transformed signal tSYM-H.}
 \end{center}
\end{figure}

 \section{Results}
 \label{sec:results}
In \cite{RR} it is shown that the fluctuation amplitude (or more precisely; the one-timestep increment) $\Delta y(t)$ of the AE index is on the average proportional to the instantaneous value $y(t)$ of the index. This gives rise to a special kind of intermittency associated with multiplicative noises, and leads to a  non-stationary  time series of increments. However, the time series $\Delta y(t)/y(t)$ is  stationary, implying that the stochastic process  $x(t)=\log y(t)$ has stationary increments. Thus, a signal  with stationary increments, which still can exhibit a multifractal intermittency,  can  be constructed by considering the logarithm of the AE index. Similar properties pertain to the SYM-H index, although in these cases we have to add a constant $c_1$ before taking the logarithm, i.e. $x(t)=\log{(c_1+c_2y(t))}$ has stationary increments. Using the procedure described in \cite{RR} the  estimated coefficients are  $c_1=0.7725$ and $c_2=0.0397$. In Figure 9a we show the increments for the original SYM-H data, while in Figure 9b we show the increments for the transformed signal $x(t)$, which in the following will be denoted tSYM-H.

\subsection{Scaling  of storm- and solar wind parameters}

In this section, we employ EMD and variogram analysis to tSYM-H, IMF $B_{z}$ and solar wind flow speed $v$. The EMD analysis is used to compute intrinsic mode functions (IF) for time intervals of $50000$ minutes using data for the entire period from January 2000  till December 2005.
The empirical  variance  estimates $E$ versus mean period $T$ for each IF component in  tSYM-H, $B_{z}$, and  $v$ are shown as   log-log plots in Figure 10a.
In section \ref{EMDmethod} we mentioned that \cite{FG04} has demonstrated that   for  fractional Gaussian noise the slope $\zeta$ is equal to $\zeta=2H-2$, where $H$ is the Hurst exponent. This estimate for the slope seems valid  for our data as well, as
is shown in the figure from  comparison with the variogram, even though the time series on scales up to $10^4$ minutes are non-stationary processes having the character of fractional Brownian motions \citep{RR}. The results from the two different methods shown in Figures 10a and 10b are roughly consistent, using the relations $h=H-1$ and  $\zeta=2H-2$, which implies $2h =\zeta$.  In practice, we have calculated $\zeta$ from EMD as a function of $2h$ for fractional Gaussian noises and motions with self-similarity
exponent $h$, and have derived a relation $\zeta=0.94\ (2h)+0.1143$.

The variogram represent a second order structure function:
\begin{equation}
 \gamma_{k}=\frac{1}{\ (N-k)}\ \sum_{n=1}^{N-t}(s_{n+k}-s_{n})^{2},  \label{eq12}
\end{equation}
which scales with a time-lag $k$ as $\gamma_{k}=k^{2h}$,  $h$ is denoted as selfsimilarity exponent, and $s$ is a time series.
Note that a Hurst exponent $H>1$ implies that the process is a nonstationary motion, and if the process is selfsimilar, the selfsimilarity exponent is $h=H-1$.  In our terminology a white noise process has Hurst exponent $H=0.5$ and a Brownian motion has $H=1.5$.

From Figure 10a we observe  three different scaling regimes for tSYM-H.  For time scales less than a few hundred minutes it scales like an uncorrelated motion ($h\approx 0.5$). On time scales from a few hours to a week it scales  as an antipersistent motion ($h\approx 0.25-0.35$ depending on analysis method), and on longer time scales than a week  it is close to a stationary pink noise ($h\approx 0$). 
Similar behavior was observed for $\log AE$ in \cite{RR}, but there the break between  non-stationary motion and stationary noise (where $h$ changes from $h>0$ to $h\approx 0$) occurs already after about 100 minutes, indicating the different time scales involved in ring current (storm) dynamics and electrojet (substorm) dynamics.

Results for $v$  indicate a regime with antipersistent motion ($h=0.25$) up to a few hundred minutes,  and then an uncorrelated or weakly persistent motion ($h=0.5$) up to a week. On longer time scales than this the variogram indicates that the process is stationary.

The exponent $h$ for  $B_{z}$ can not be estimated from the variogram since it is difficult to obtain a linear fit  to the concave curve in Figure 10b.  The  concavity is less pronounced in the curve derived from the EMD method in Figure 10a and  $\zeta=0.47$, corresponding to an antipersistent motion ($h=0.23$), can be  estimated on time scales up to a few hundred minutes. $B_{z}$ becomes stationary already after a few hundred minutes, which is similar to the behavior in $\log AE$, as pointed out by \cite{RR}. In \cite{RR2} the concavity of the variogram follows from modelling $B_z$ as a (multifractal) Ornstein-Uhlenbeck process with a strong damping term that confines the motion on time scales longer than 100 minutes. Accounting for this  confinement the ``true" selfsimilarity exponent of the stochastic term turns out to be  $h\approx 0.5$. Thus the antipersistence derived from the EMD analysis may be a spurious effect from this confinement. The conclusion in \cite{RR2} is that  $B_{z}$ and $\log AE$ behave
as uncorrelated motions up to the scales of a few hours and become stationary on scales longer than this. Moreover, the stochastic term modelling the two signals share the same multifractal spectrum. In comparison, tSYM-H and $v$ are non-stationary motions on scales up to a week before they reach the stationary regime.

\subsection{Change of determinism during storm times}\label{stormdeterminism}

In Figure 11 we show $L_n$ and $\Lambda(\tau)$ for tSYM-H and its surrogate time-series with randomized phases of Fourier coefficients. We observe that $L_n$ and $\Lambda(\tau)$ for the surrogate time series does not deviate from those computed from the original tSYM-H, indicating that the dynamics of
tSYM-H is not low-dimensional and nonlinear. The same results are obtained for IMF $B_{z}$ and flow speed $v$ (not shown here).

\begin{figure} \label{fig10}
\begin{center}
 \includegraphics[width=8cm]{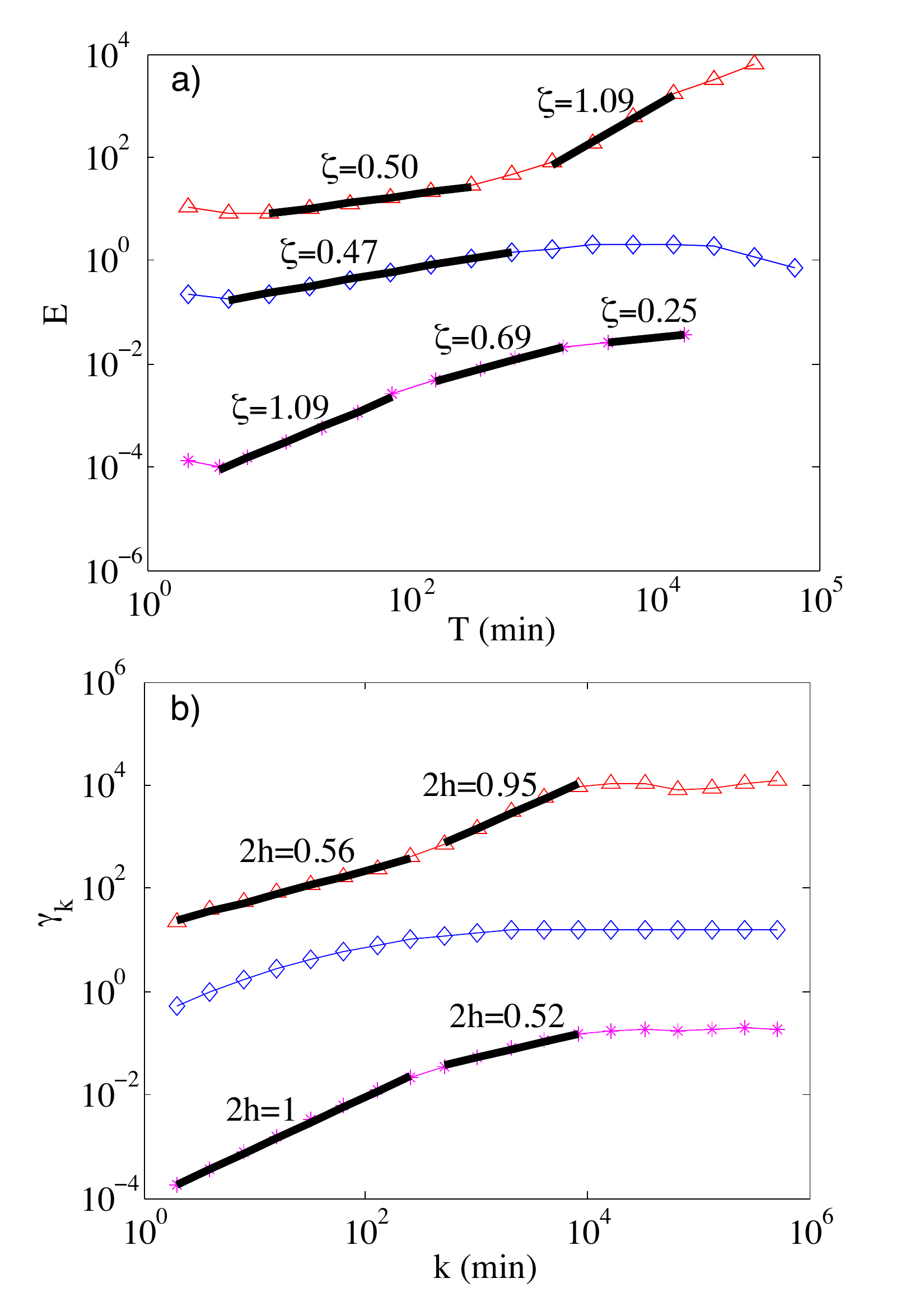}
 \caption{a) The empirical  variance  estimates $E$ versus mean period $T$ for each IF component in  tSYM-H, $B_{z}$, and  $v$  shown as   log-log plots. b) The variogram $\gamma_{k}$ shown in log-log plot. In both panels stars are  for tSYM-H, diamonds  for  IMF $B_{z}$, and triangles for $v$. Note that a generalization of  the result $\zeta =2H-2$ in \cite{FG04} yields $2h=\zeta$. }
 \end{center}
 \end{figure}

\begin{figure} \label{fig11}
\begin{center}
\includegraphics[width=8cm]{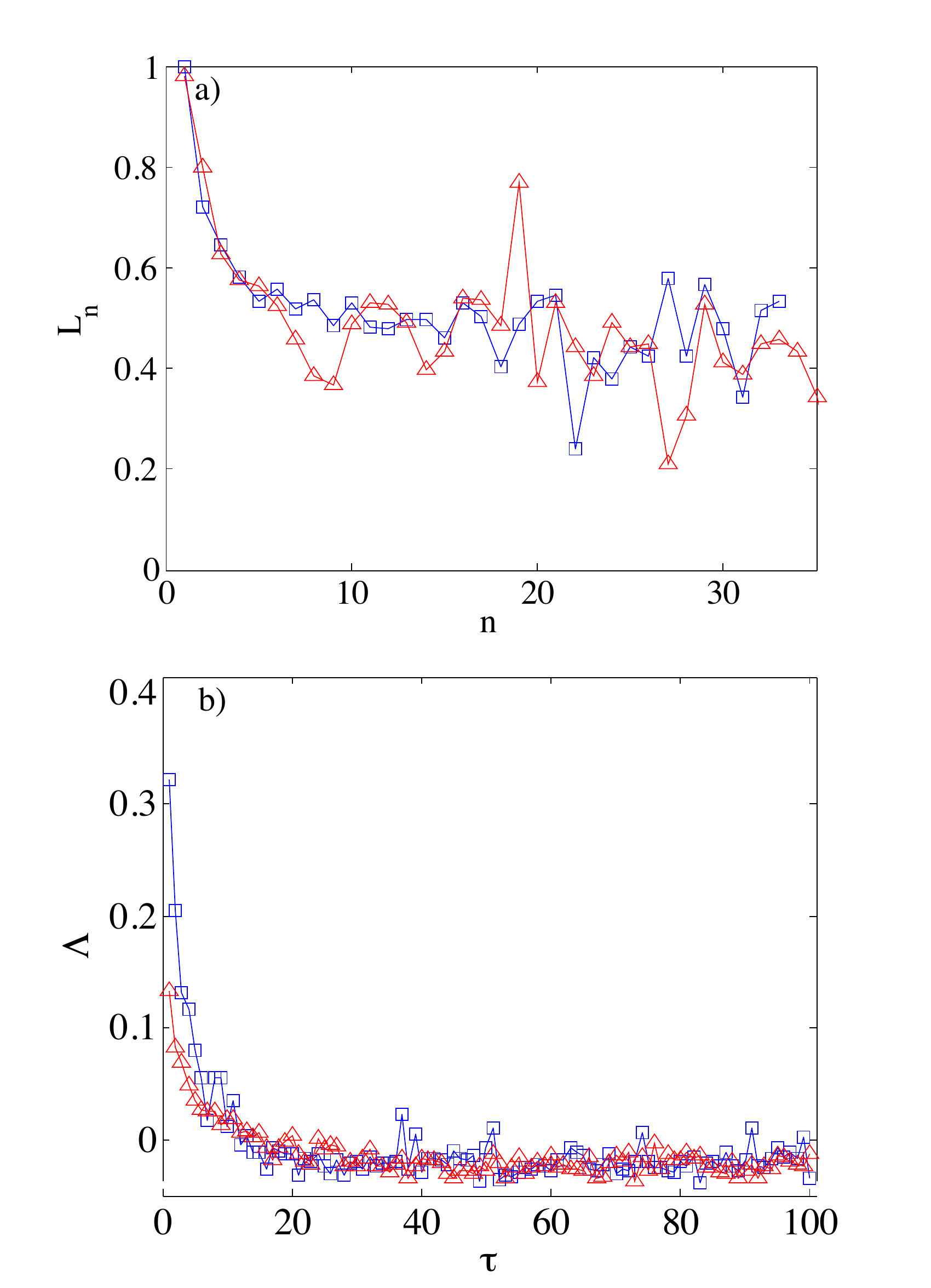}
\caption{ a) $L_n$: square symbols are for tSYM-H, and  triangles are for this signal after randomization of phases of Fourier coefficients. b) $\Lambda(\tau)$: squares is for tSYM-H, and  triangles are after randomization of  phases.}
\end{center}
\end{figure}

\begin{figure} \label{fig12}
\begin{center}
 \includegraphics[width=8cm]{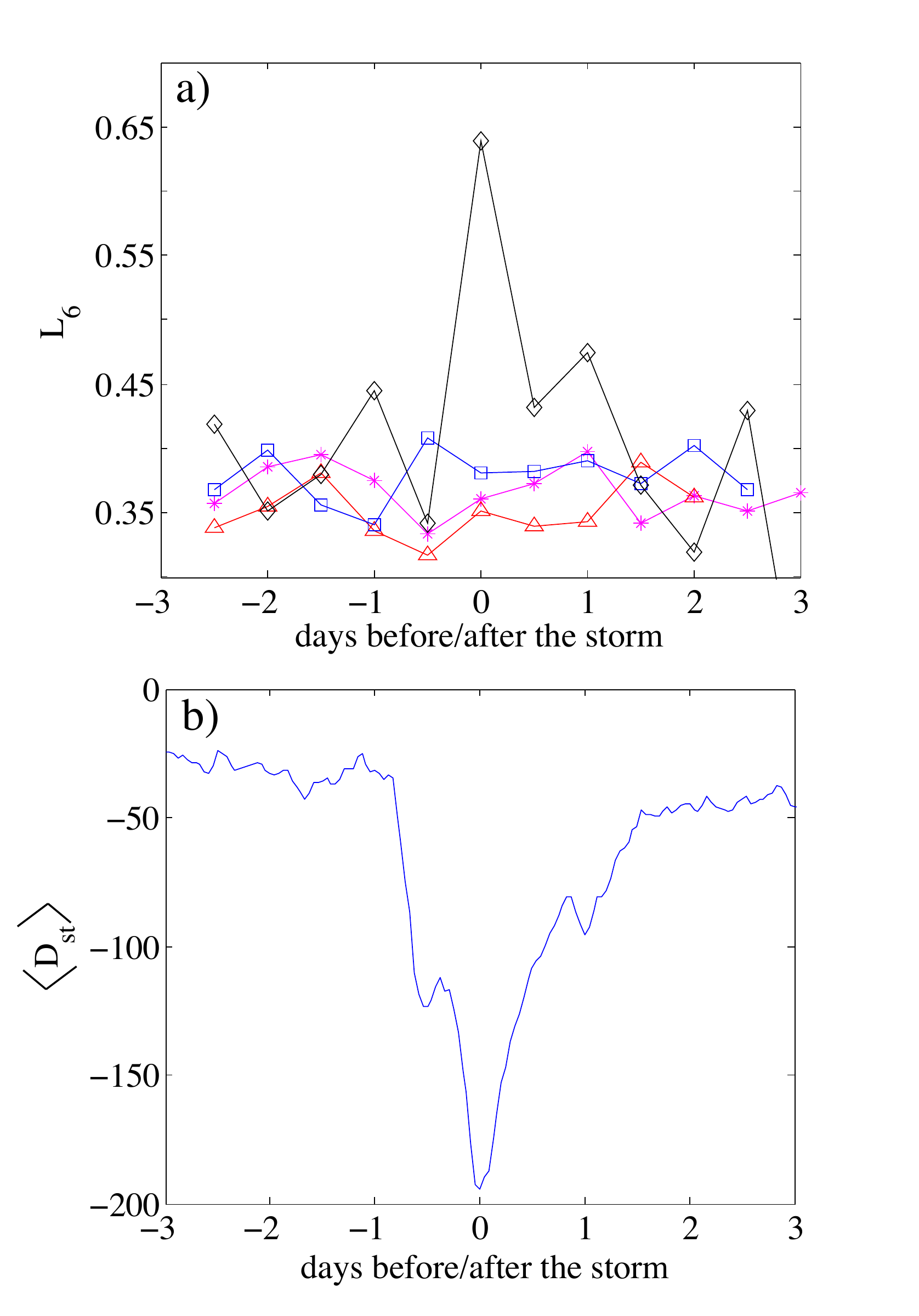}
 \caption{a) $L_{6}$ for IMF $B_{z}$ (squares), $v$ (triangles), tSYM-H (diamonds), fO-U (stars) computed before, during and after storm onset. The values for $L_6$ are computed using 12 hour intervals and are averaged over ten different storms. b) The $D_{st}$ index averaged over the ten storms.  }
\end{center}
\end{figure}

\begin{figure} \label{fig13}
\begin{center}
 \includegraphics[width=8cm]{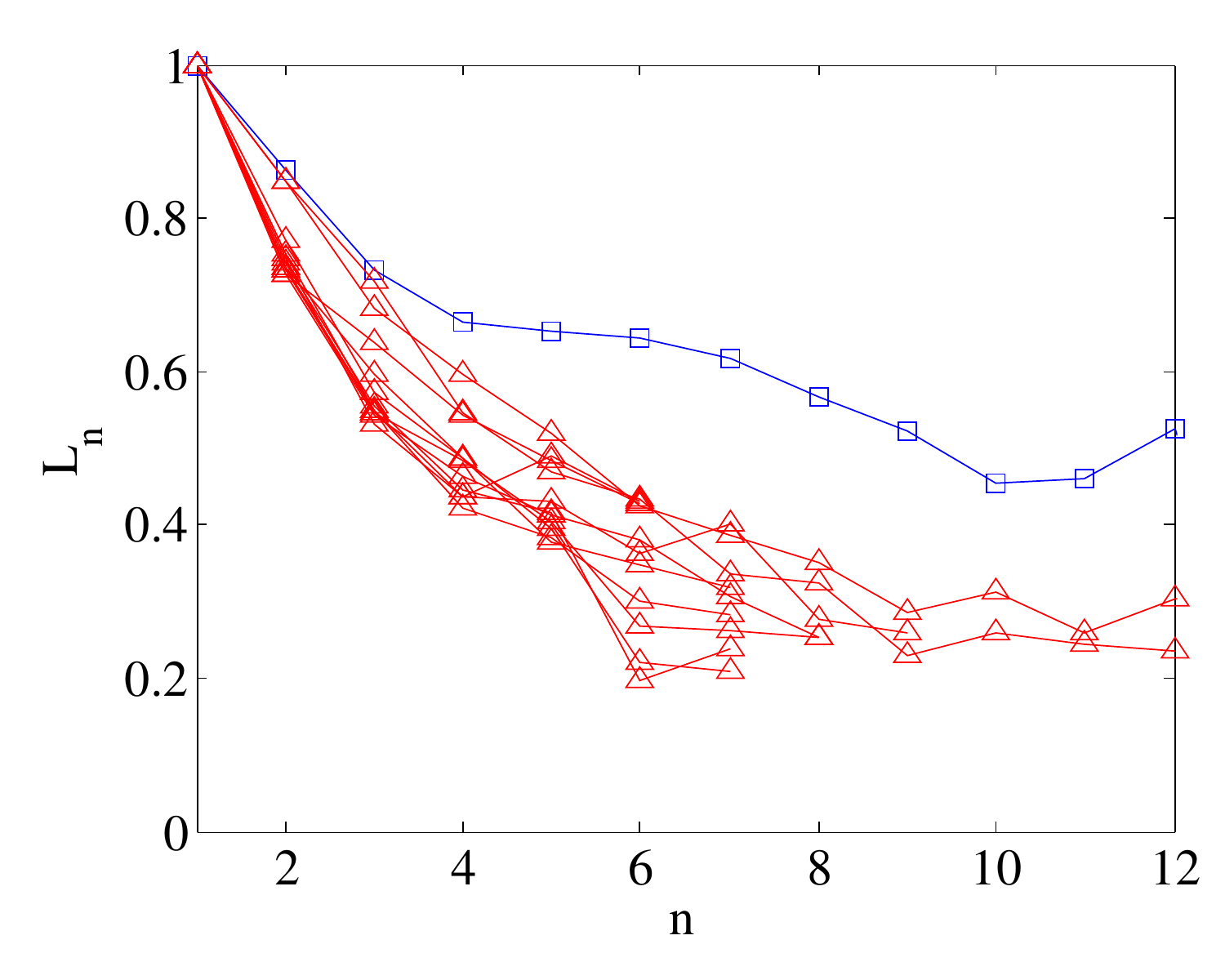}
 \caption{$L_{n}$ for tSYM-H, where the triangles are the mean of $L_{n}$ computed  3 days before and after the storm for ten different storms. These curves represent non-storm conditions. The upper curve (squares) is the mean over all ten storms computed at the time of the
 $D_{st}$ minimum, i.e. it represents the $L_n$-curve  around storm onset. Many curves are terminated for $n_{max}<12$ because there were no boxes  with more than $n_{max}$ passages of the phase-space trajectory.}
\end{center}
\end{figure}

\begin{figure} \label{fig14}
\begin{center}
 \includegraphics[width=8cm]{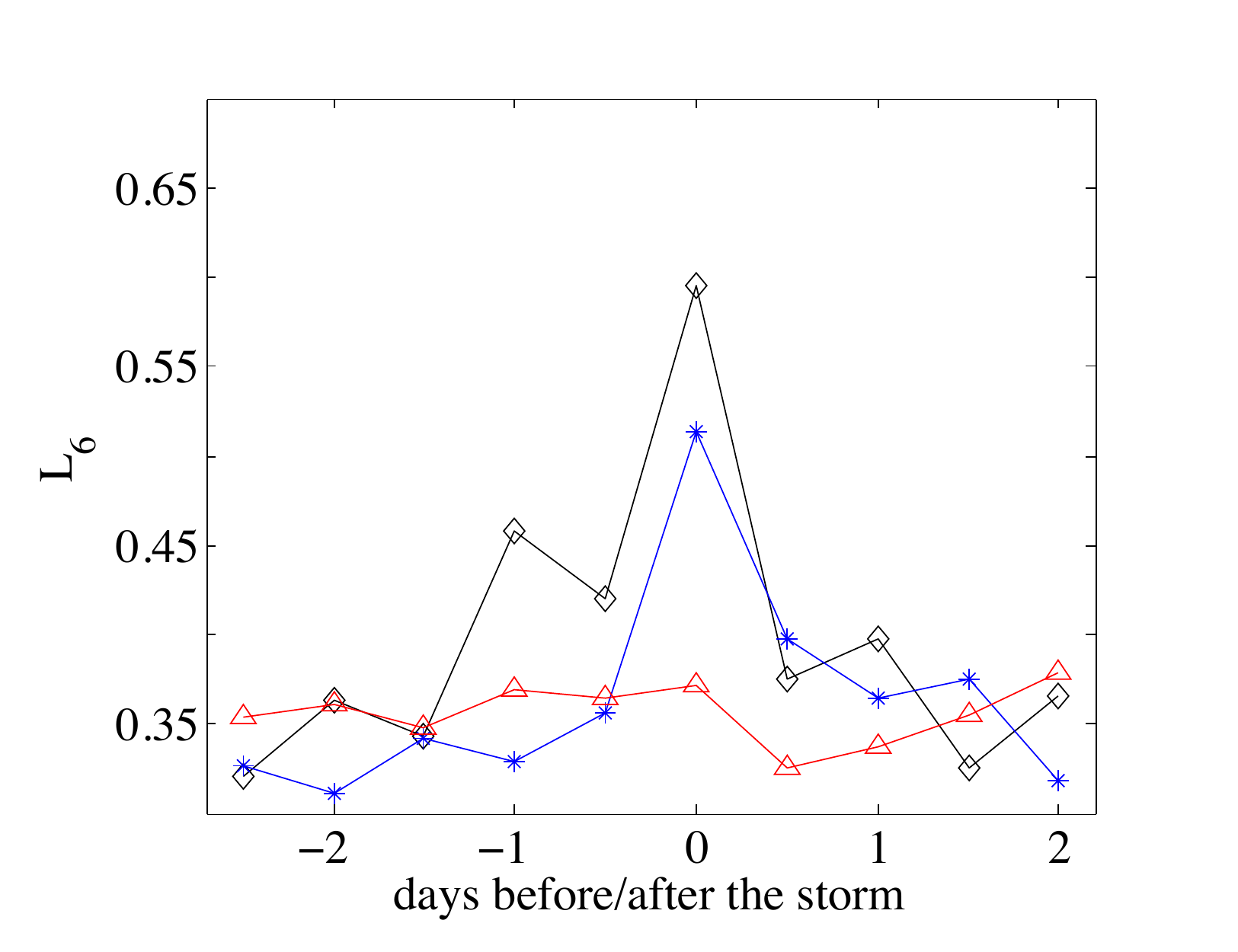}
 \caption{$L_{6}$ averaged over ten storms for time series where missing points have been interpolated; tSYM-H (diamonds), tSYM-H$^{\star}$ (stars), $\sqrt(P_{dyn})$ (triangles)}
\end{center}
\end{figure}

In the following analysis we  test for determinism in tSYM-H, $B_{z}$ and  $v$ for ten intense storms. The reference point in our analysis is the storm's main phase, and then we analyze all the data spanning the time interval three days before and three days after the storm in tSYM-H,  $B_z$ and $v$.
We compute  $L_{n}$ for $n=6$ with a time resolution of 12 hours. The choice of $n=6$  from the $L_n$ -curve is a compromise between clear separation between low-dimensional and stochastic dynamics and small error bars (which increase with increasing $n$). In order to improve statistics for $L_{6}$, we compute determinism using data from all ten storms. This means that each $L_{6}$ is computed over 12 hours interval over 10 storms, which gives $12\cdot 60\cdot 10=7200$ points.
As a reference, we compute $L_{6}$ for the fO-U process, whose coefficients are fitted by the least square regression to the SYM-H index during investigated storms. 
In all computations, we use embedding dimension $m=7$, time-delay $\tau=20$, and $b=1$.
In Figure 12a we plot $L_{6}$ for  tSYM-H   $B_{z}$,  $v$, and  fO-U, and
in Figure 12b   the  $D_{st}$ index averaged over all ten storms is plotted, since this index shows precisely when the storm takes place. We can observe  that $L_{6}$ is essentially the same for $B_{z}$, $v$ and fO-U, and stays approximately constant during the course of a storm. However, $L_{6}$ for tSYM-H increases during  storm time. In order to demonstrate that the change in $L_{6}$ is significant, we plot in Figure 13  $L_{n}$ for tSYM-H, where the triangles are the mean of $L_{n}$ computed  3 days before and after the storm for ten different storms. These curves represent non-storm conditions. The upper curve (squares) is the mean over all ten storms computed at the time of the
 $D_{st}$ minimum, i.e. it represents the $L_n$-curve  around storm onset.

In addition, we test determinism for the quantity SYM-H$^{\star}$=0.77 SYM-H-11.9$\sqrt{P_{dyn}}$ \citep{KL2003}, where $P_{dyn}$ is the Solar wind's dynamic pressure. SYM-H$^{\star}$ is a corrected index where the effect of the magnetopause current due to $P_{dyn}$ is subtracted, and thus represents the ring-current contribution to SYM-H.
In order to obtain stationary increments we analyze a transformed index;  tSYM-H$^{\star}$=$\log$($c_1$+$c_2$SYM-H$^{\star}$), where $c_1=1.7694$, and $c_2=0.0292$.
Since some data points are missing in $P_{dyn}$, we have made a  linear interpolation over the missing points. It seems that the interpolation decreases determinism in  tSYM-H$^{\star}$, and for  reference we compute determinism for the interpolated tSYM-H, where the interpolation is done over the same points as in tSYM-H$^{\star}$, even though tSYM-H does not have missing points. In Figure 14 we plot $L_{6}$ for tSYM-H$^{\star}$, tSYM-H,  and $\sqrt(P_{dyn})$. We see that the determinism in tSYM-H$^{\star}$ is lower than that in tSYM-H, but it  still increases during the storm. On the other hand $\sqrt(P_{dyn})$ shows no change in determinism during storm events.

\subsection{Discussion of results on determinism}

The determinism (as measured by $L_6$)  of the storm index tSYM-H and tSYM-H$^{\star}$ has been shown to exhibit a pronounced increase at storm time. A rather trivial explanation of this enhancement would be that it is caused by the ``trend" incurred by the wedge-shaped drop and recovery of the storm indices associated with a magnetic storm. We test this hypothesis by superposing such a wedge-shaped pulse  to an fO-U process and compute $L_6$.  
Next, we take tSYM-H for ten storms and for each set of data subtract the wedge-shaped pulse (computed by a moving-average smoothing). The residual signal represents the ``detrended" fluctuations.
The result is shown in Figure 15 and reveals that the trend in fO-U process has no discernible influence on the determinism during the storm while, on the other hand, we observe that  the enhancement of $L_{6}$  around storm time persists in the ``detrended" fluctuations. This result suggests that the increasing determinism during storms is a result of an enhanced low-dimensional component in the storm indices. 
As mentioned in section \ref{determinism} for low-dimensional dynamics, nonlinearity may be important for the measure of determinism. For a nonlinear, low-dimensional system the destruction of nonlinear coupling by randomizing phases of Fourier coefficients will in general reduce the determinism, while for a linear, stochastic process we will observe no such effect. But what role will nonlinearity play if it is introduced in the deterministic terms of a stochastic equation? The deterministic term in the fractional Langevin equation representing the fO-U is a linear damping term. However, the best representation of the damping/drift term in an fO-U model for tSYM-H is not linear. Following \cite{RR2}, if $y=$tSYM-H, the drift term is given as the conditional probability density $\delta y(t,\delta t)$ given that $y(t)=y$:
\begin{equation} 
M(y,\delta t)=E\lbrack \delta y(t,\delta t)\mid y(t)=y\rbrack.
\end{equation}
In fO-U $M(y,\delta t)$ is a linear function of $y$, but a polynomial  fit to drift term derived from tSYM-H data requires a sixth order polynomial, confirming the nonlinearity of the tSYM-H process. This is shown in Figure 16. 
Next, we test determinism for the nonlinear fO-U process, whose scaling exponent $h$ is estimated from the variogram of tSYM-H, and where $m=8$ and $\tau=10$ is used.
Figure 17a shows  $L_{n}$   for numerical realizations of this process compared with the same analysis after randomization of the phases of the Fourier coefficients. The result reveals that the nonlinear fO-U process is not more deterministic than its randomized version. Next, we form the composite time series $x=x_{L}+1.85x_{n}$, where $x_{L}$ is the solution of the Lorenz system and $x_{n}$ is the nonlinear fO-U process, both signals with  zero mean and unit  variance. Again, embedding dimension $m=8$ is used. Now the  $L_{n}$ -curve is lowered when the phases are randomized, as shown in Figure 17b,   which confirms our conjecture that determinism is a measure of low-dimensionality.

\begin{figure} \label{fig15}
\begin{center}
 \includegraphics[width=8cm]{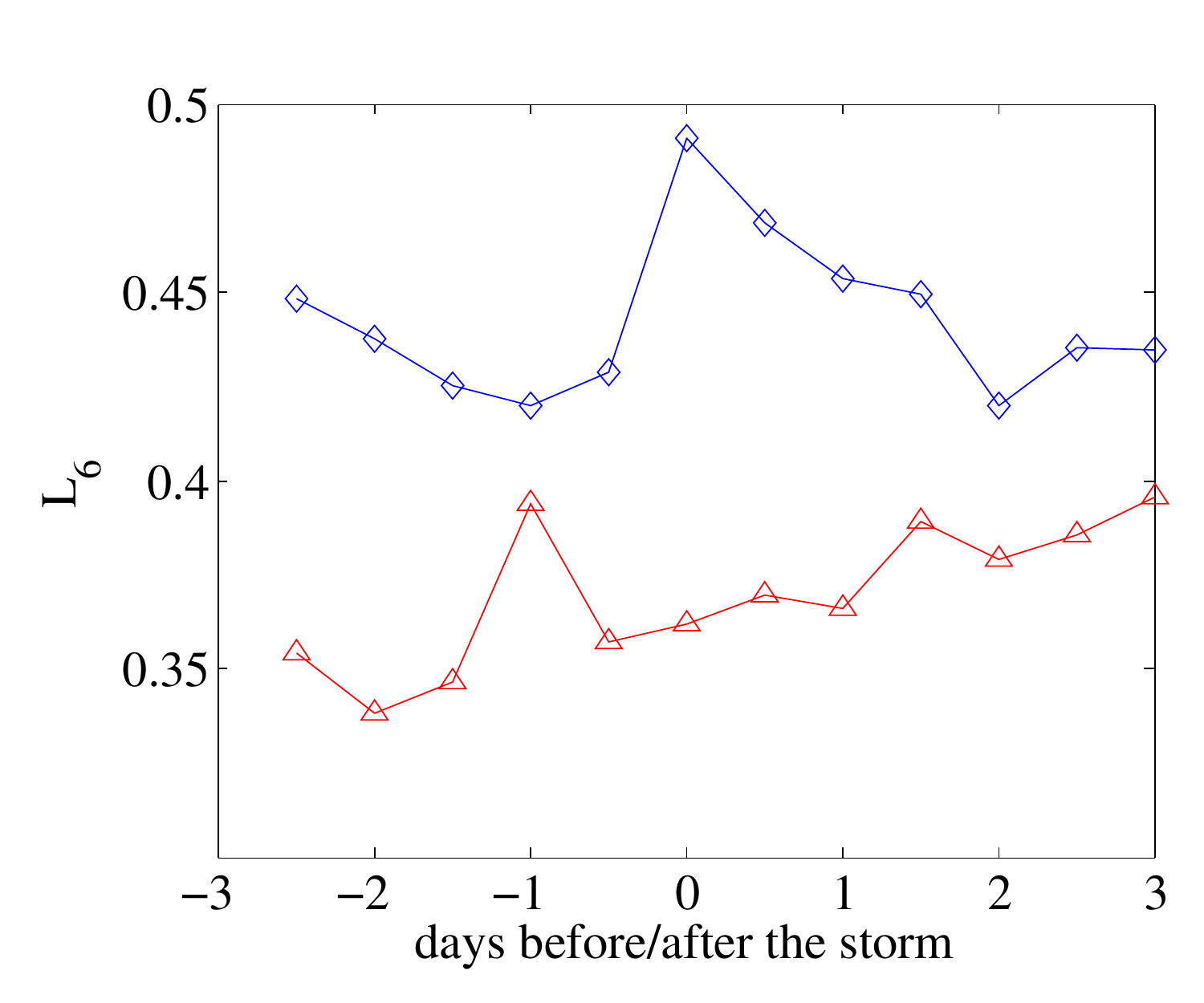}
 \caption{$L_{6}$: triangles are derived from an fO-U process with a ``storm trend'' imposed, diamonds are derived from the ``detrended" tSYM-H.}
\end{center}
\end{figure}
\begin{figure} \label{fig16}
\begin{center}
 \includegraphics[width=8cm]{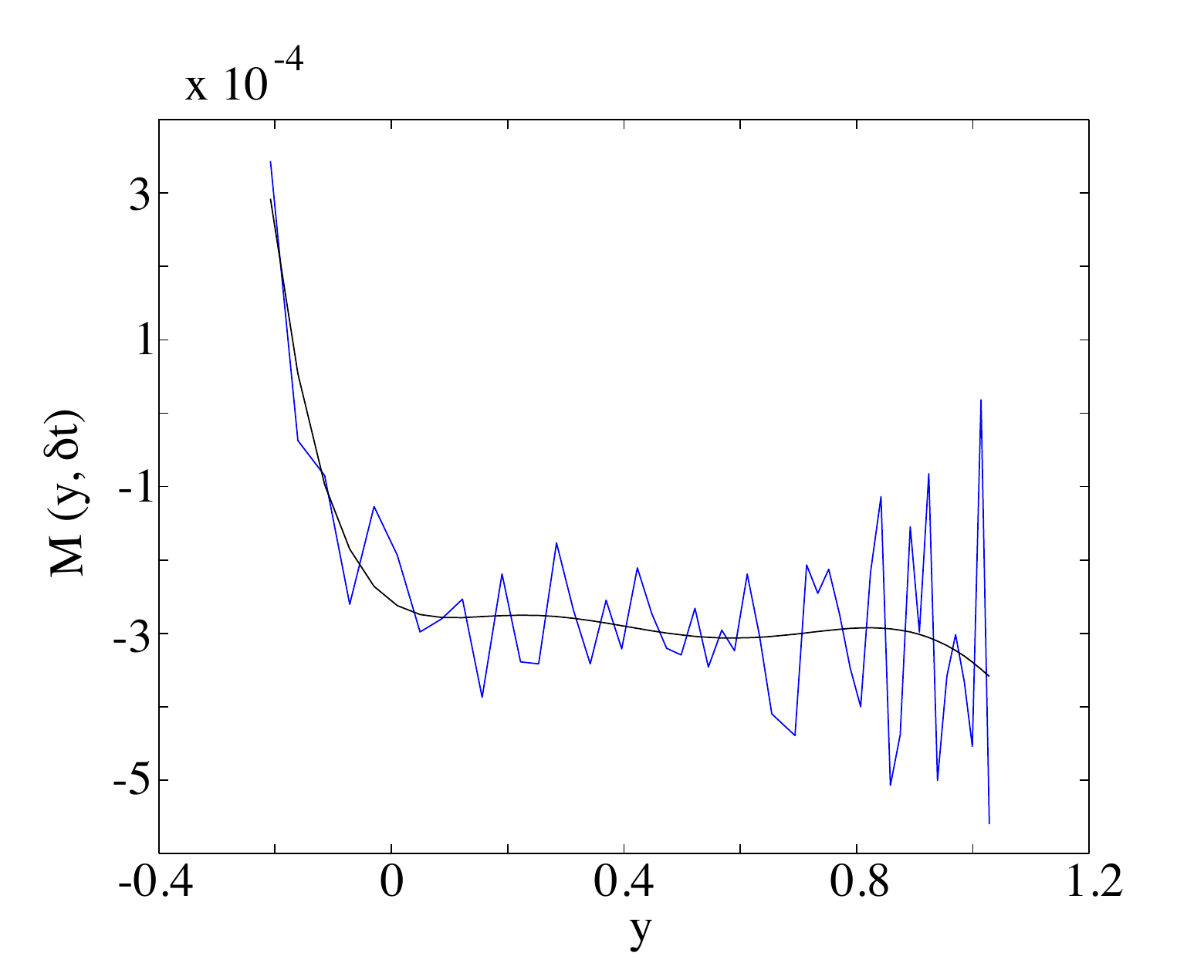}
 \caption{The drift term in the fO-U equation computed from tSYM-H. The smooth solid curve is  a six-order polynomial fit.}
\end{center}
\end{figure}
\begin{figure} \label{fig17}
\begin{center}
 \includegraphics[width=8cm]{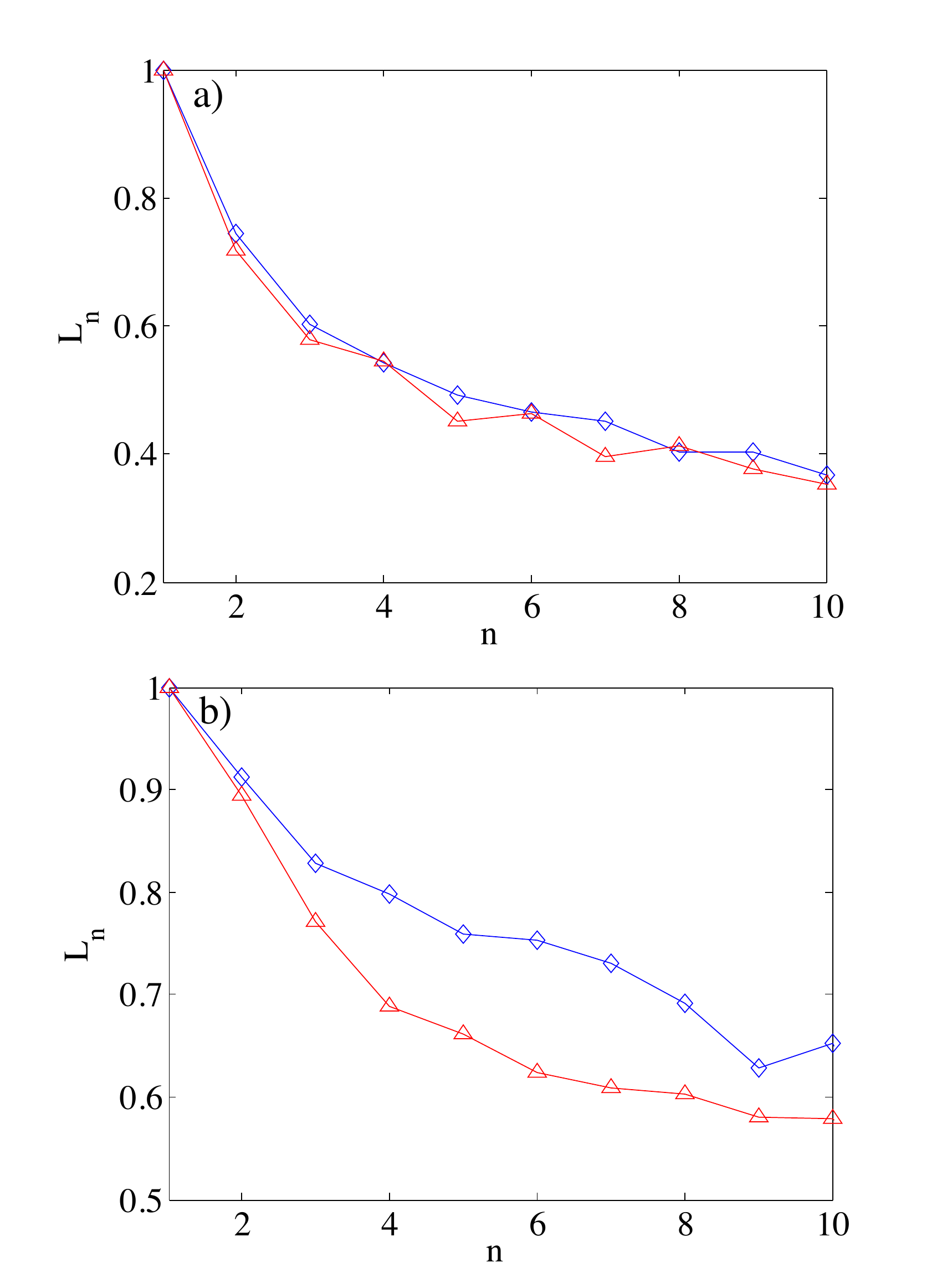}
 \caption{$L_{n}$. a) diamonds are derived from numerical solutions of the nonlinear fO-U. Triangles are  from these solutions after randomization of phases of Fourier coefficients. b) diamonds are derived from numerical solutions of the nonlinear fO-U with a solution of the  {\it x} component of the Lorenz system superposed.  Triangles are  from the latter signals after randomization of phases.}
\end{center}
\end{figure}
\begin{figure} \label{fig18}
\begin{center}
 \includegraphics[width=8cm]{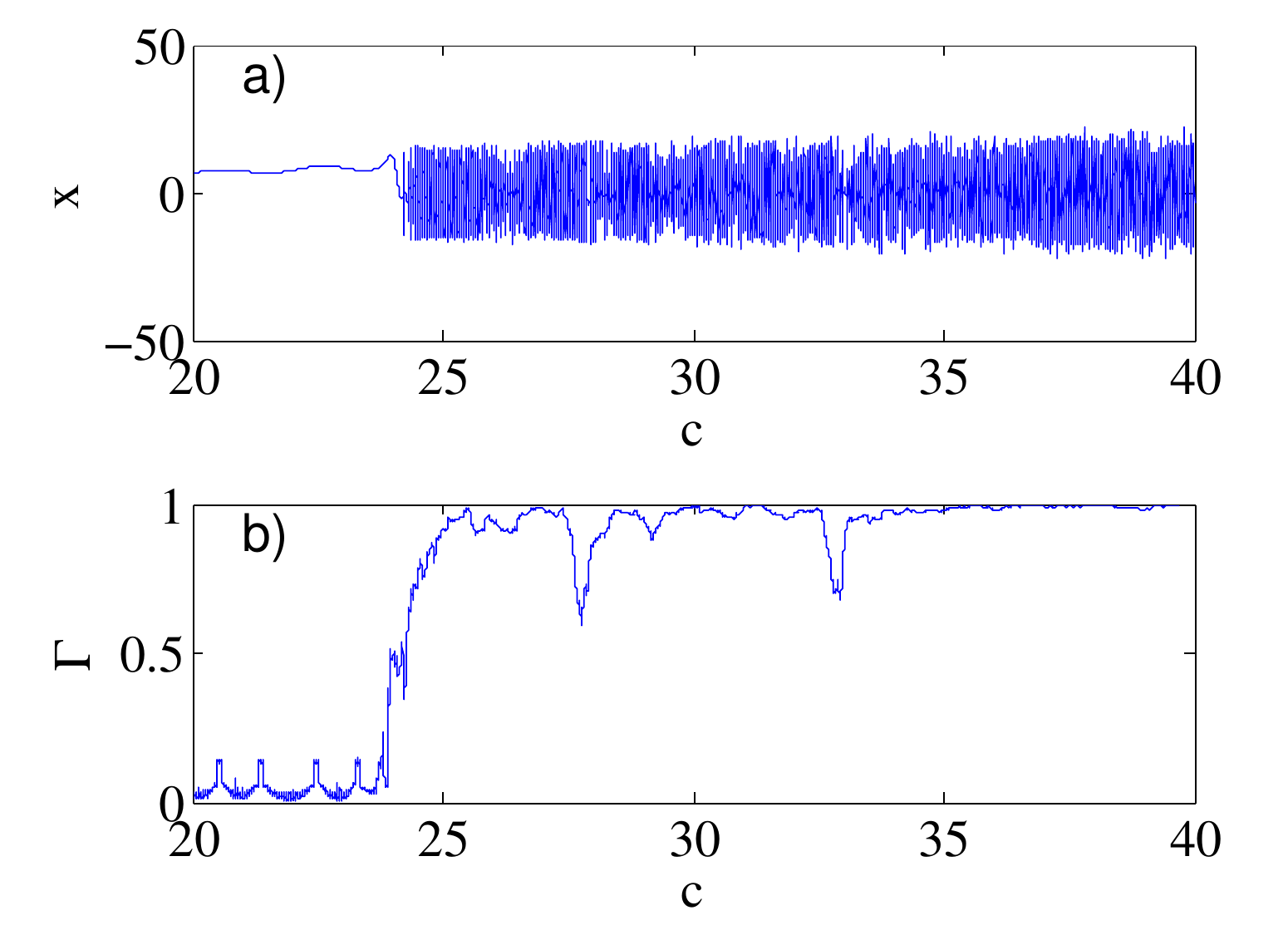}
 \caption{a) {\it x} component of the Lorenz system as a function of the parameter {\it c}. b) $\Gamma \equiv \langle l^{-1} \rangle$ as a function of the parameter {\it c}.}
\end{center}
\end{figure}

\begin{figure} \label{fig19}
\begin{center}
\includegraphics[width=8cm]{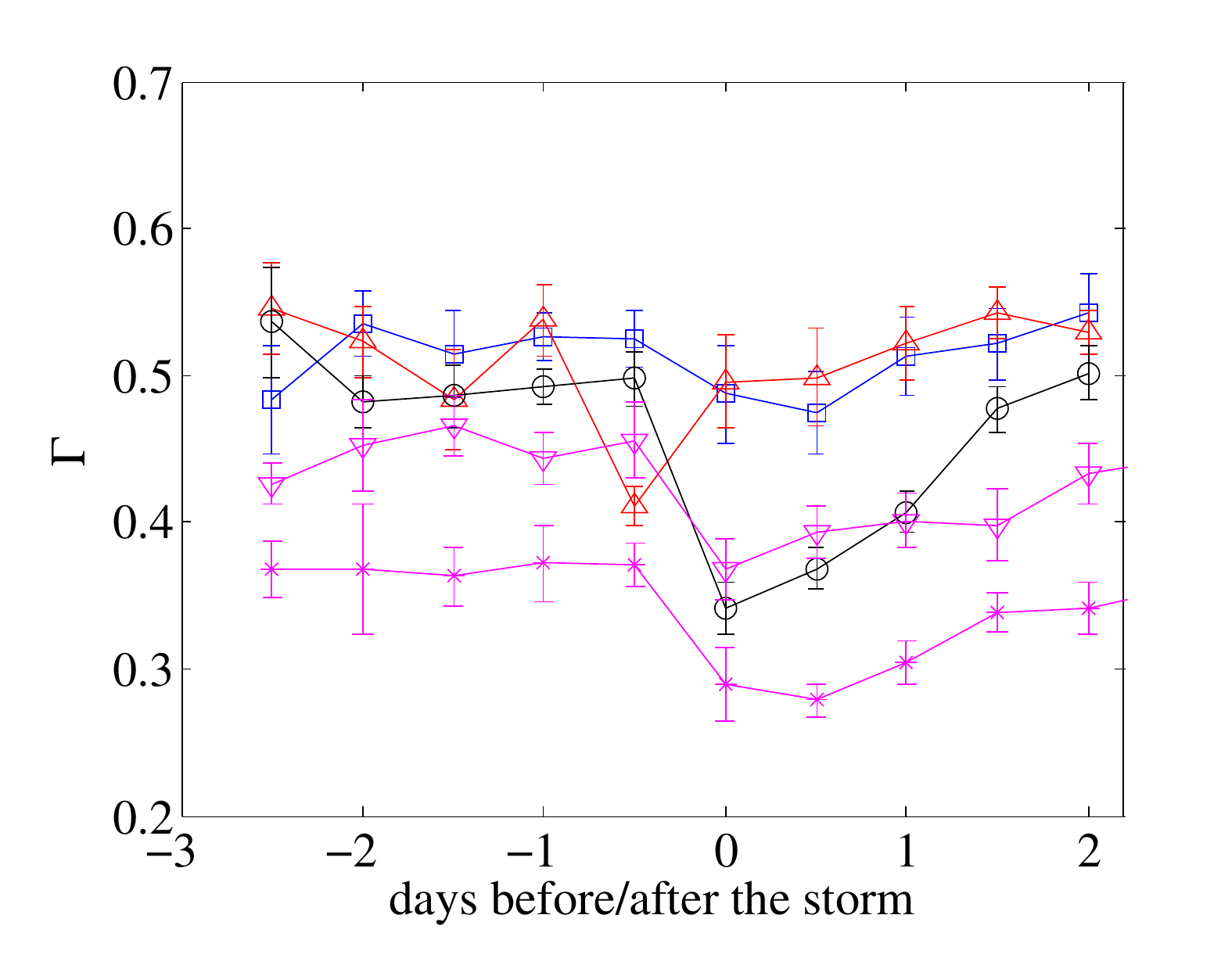}
 \caption{a) $\Gamma$ for tSYM-H (stars), tSYM-H$^{\star}$ (circles), $B_{z}$ (squares), $v$ (upward triangles), detrended tSYM-H (downward triangles) averaged over ten storms. Error bars represent standard deviation based on data from these ten storms. Time origin is defined by the minimum of the average $D_{st}$ index for the ten storms.}
\end{center}
\end{figure}
\begin{figure} \label{fig20}
\begin{center}
\includegraphics[width=8cm]{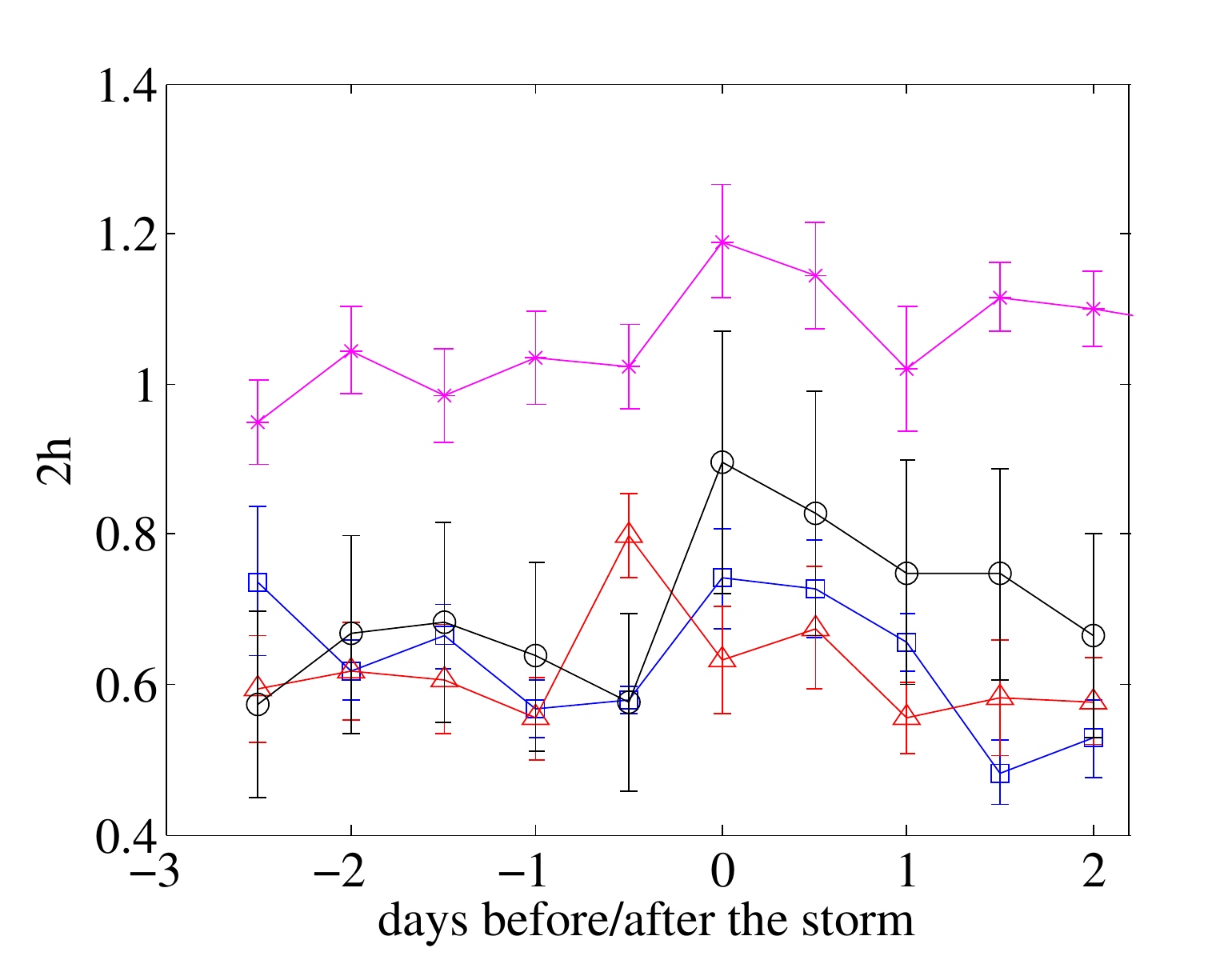}
 \caption{a) $2h$ for  tSYM-H (stars), tSYM-H$^{\star}$ (circles), $B_{z}$ (squares), $v$ (triangles) averaged over the same storms as in the previous figure.}
\end{center}
\end{figure}
\begin{figure} \label{fig21}
\begin{center}
 \includegraphics[width=8cm]{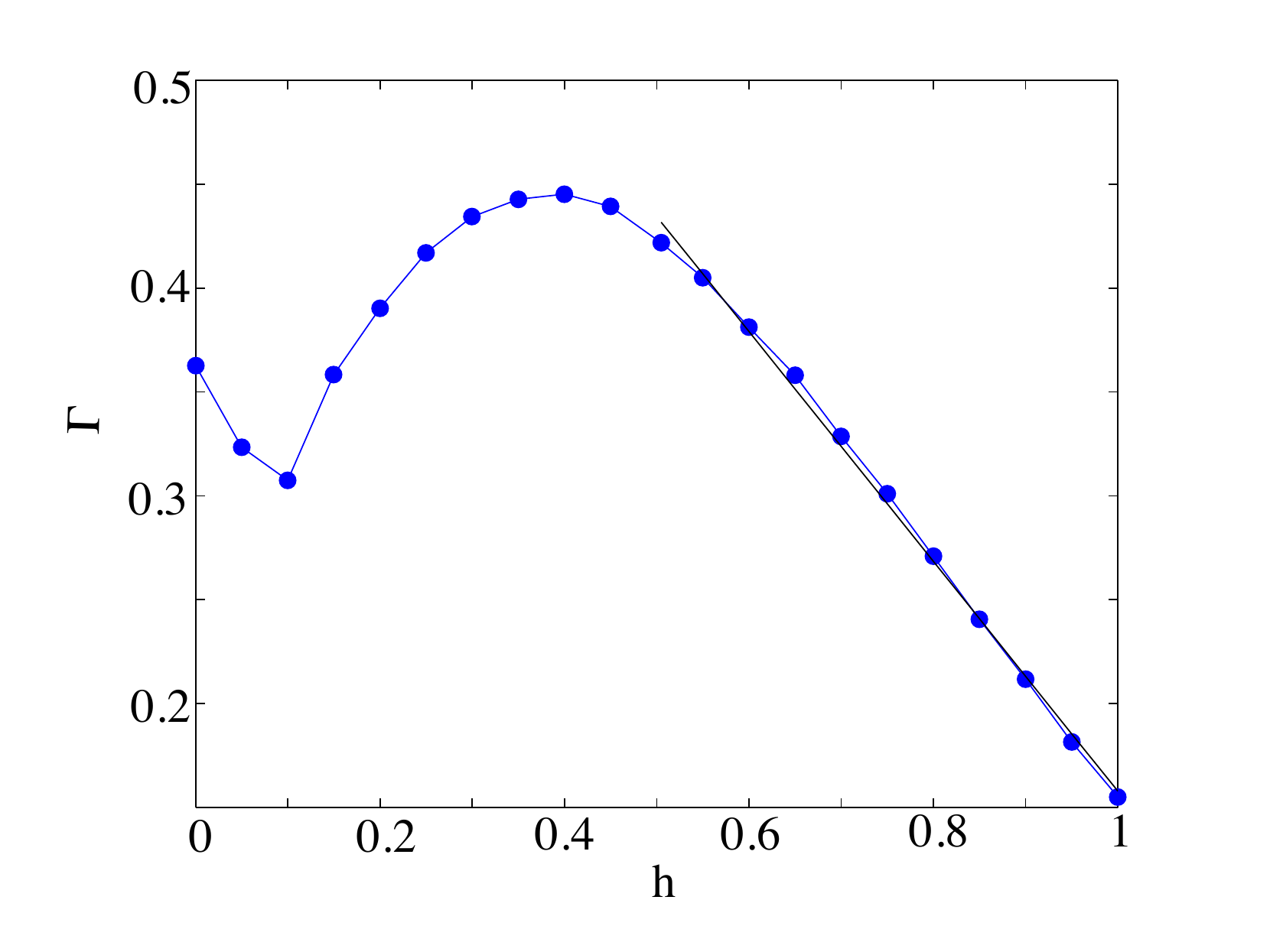}
 \caption{$\Gamma$ vs. $h$ computed from numerical realizations of the fO-U process.}
\end{center}
\end{figure}
\subsection{Change of predictability during magnetic storms}
\label{Sec:predictabilityresults}
Even though we deal with a predominantly stochastic system, its correlation and the degree of predictability changes in time, and our hypothesis is that  abrupt transitions in the dynamics take place during events like magnetic storms and substorms. We therefore employ recurrence plot quantification analysis as  a tool for detection of these transitions. 

We compute the average inverse diagonal line length $\Gamma\equiv \langle l^{-1}\rangle$ as defined in equation (\ref{eq4}), but 
the same results can be drawn from other quantities that can be derived from the recurrence plot \citep{RQA}.
$\Gamma$ can be used as a proxy for the positive Lyapunov exponent in a system with  chaotic dynamics, and is sensitive to the transition  from regular to chaotic behavior, as can be shown heuristically for the case of the Lorenz system, where we use $a=10$, $b=8/3$, and $c$ is varied from 20 to 40, such that transient behavior is obtained. For $c=24.74$ a Hopf bifurcation occurs, which corresponds to the onset of chaotic flows. 
In Figure 18a  we plot a bifurcation diagram for the {\it x} component of the Lorenz system as a function of the parameter $c$,  while in Figure 18b we show $\Gamma$ for the {\it x} component again as a function of the parameter {\it c}.
Similar results have been obtained from the longest diagonal length, when applied to the logistic map \citep{Trulla96}.

In the following analysis, we use embedding dimension $m=1$, because  the results do not seem dependent on $m$ and because,  in the case of  stochastic or high-dimensional dynamics, a topological embedding  cannot be achieved for any reasonable embedding dimension. This fact demonstrates the 
robustness of the recurrence-plot analysis, which responds to changes in the dynamics of the system even if it is a stochastic or high-dimensional system for which no proper phase-space reconstruction is possible. 

Since reduction in $\Gamma$ means  increase of predictability it may  also be a signature of higher persistence in a stochastic signal. This motivates plotting 
$\Gamma$ and $2h$ (computed as a linear fit from the variogram over the time scales up to 12 hours) for solar wind parameters and magnetic indices.
Figure 19  shows $\Gamma$ for tSYM-H, $B_{z}$, $v$, tSYM-H$^{\star}$ and detrended tSYM-H averaged over 10 magnetic storms. Figure 20 shows  the same for $2h$, but detrended tSYM-H is not shown since
its $2h$ changes insignificantly during the course of the storm.
We observe that the increase in the predictability and persistence does not occur simultaneously for all observables. While $B_{z}$, tSYM-H, detrended tSYM-H and tSYM-H$^{\star}$ get the most predictable during or after the main phase of the storm, solar wind's flow speed becomes the most predictable {\em prior} to the storm's main phase.
From a hundred realizations of the fO-U process  generated numerically with the coefficients in the stochastic equation fitted to model the  tSYM-H signal, we find $\Gamma=0.4$ and $2h=1$, in good agreement with the results obtained  from the tSYM-H time series.  
The general relationship between $\Gamma $ and $h$ can also be explored through numerical realizations of   fO-U processes. Figure 21 shows $\Gamma$ computed for varying $h$ as a mean value of  100 realizations of such a process for each $h$. For  persistent motions ($h>0.5$) there is a linear dependence between  $\Gamma$ and $h$, and a best fit  yields 
\begin{equation}
\Gamma\approx 0.72-0.57 h. \label{gamma-h}
\end{equation}
This analysis shows the importance of  $\Gamma$ as a universal measure for predictability: in low-dimensional systems it is a proxy for the Lyapunov exponent, while for persistent stochastic motions it is a measure of
persistence through equation (\ref{gamma-h}).

\section{Conclusions}
The storm index SYM-H and the solar wind observables (flow velocity $v$ and  IMF $B_{z}$) show no clear signatures of low-dimensional dynamics during quiet periods. However, low-dimensionality increases in SYM-H and SYM-H$^{\star}$ during storm times, indicating that
self-organization of the magnetosphere takes place during magnetic storms. This conclusion is drawn from the study of ten intense, magnetic storms in the period from 2000-2003. 
Even though our analysis shows no discernible change in determinism during magnetic storms for solar wind parameters, there is an enhancement of the predictability of the solar wind observables as well as the geomagnetic storm indices during major storms. We interpret this as an increase in the persistence of the stochastic components of the signals. 
The increased persistence in the solar wind flow $v$, prior to the storm's main phase could indicate that $v$ is more important driver than $B_{z}$ during magnetic storms. This is consistent with a reexamination of the solar wind-magnetosphere coupling functions done by \cite{N2006}, who found that the most optimal function is of the form $v^{2}B\sin^{4} (\theta/2)^{2/3} $, where $\theta=\arctan(B_y/B_z)$. Also, it has been shown in \cite{P2007} through  numerical simulation, that increased $v$ changes the magnetospheric response from a steadily convecting state to highly variable in both space and time.

It has been shown in \cite{N2001} that the plasma sheet is the dominant source for the ring current based on the similarity in composition of the inner plasma sheet and ring current regions. During the main phase of the storm, ions from the plasma sheet are flowing to the inner magnetosphere on the open drift paths and then move to the dayside magnetopause. In this storm phase the ring current is highly asymmetric, as was experimentally shown by  energetic neutral atom imaging (see \cite{KL2003} and references therein). During the recovery phase, ions from the plasma sheet are trapped on  closed drift paths, and form the symmetric ring current. Therefore, the increase in determinism of the ring-current (SYM-H$^{\star}$) during storms implies increased determinism in the plasma sheet as well. 

A magnetic storm is a coherent global phenomenon investing a vast region of the inner magnetosphere, and implying large scale correlation. The counterpart of this increase of coherence is the reduction of the spontaneous incoherent short time scale fluctuations. Consequently, one should expect a reduction of the free degrees of freedom which implies an increase of determinism, i.e. the possible emergence of a low-dimensionality. 

Analysis of predictability shows significant differences between $B_z$  on one hand, and $v$ and SYM-H on the other. While the former is a non-stationary, slightly anti-persistent motion up to time scales of approximately 100 minutes, and a pink noise on longer time scales, the latter are slightly persistent motions on scales up to several days and noises on longer time scales. These differences indicate the different role the solar wind $B_z$ and the velocity $v$ play in driving the substorm and storm current systems; $B_z$ is important in substorm dynamics which  will be studied in a separate paper,  while $v$ is a major driver of storms. 

\begin{acknowledgments}
Recurrence plot and its quantities are computed by means of the Matlab package downloaded from \begin{verbatim} http://www.agnld.uni-potsdam.de/~marwan/toolbox/. \end{verbatim}
The authors acknowledge illuminating discussions with M. Rypdal and B. Kozelov. The authors would like to thank the Kyoto World Data Center for $D_{st}$ and SYM-H index, and CDAweb for allowing access to the plasma and magnetic field data
of the OMNI source. Also, comments of two anonymous referees are highly appreciated. 
\end{acknowledgments}

------------------------------- 

\end{article}
\end{document}